\documentclass[11pt,oneside,reqno]{amsart}
\usepackage[T1]{fontenc}
\usepackage[latin9]{inputenc}
\usepackage{float}
\usepackage{varwidth}
\usepackage{amstext}
\usepackage{amsthm}
\usepackage{amssymb}
\usepackage{geometry}
\usepackage{rotfloat}
\usepackage{setspace}
\makeatletter

\newcommand{\lyxmathsym}[1]{\ifmmode\begingroup\def\b@ld{bold}
  \text{\ifx\math@version\b@ld\bfseries\fi#1}\endgroup\else#1\fi}

\providecommand{\tabularnewline}{\\}
\newenvironment{cellvarwidth}[1][t]
    {\begin{varwidth}[#1]{\linewidth}}
    {\@finalstrut\@arstrutbox\end{varwidth}}

\usepackage{amsfonts}
\usepackage{amstext}

\newtheorem*{asm}{Assumption}\theoremstyle{definition}
\newtheorem{rem}{Remark}\newtheorem{ex}{Example}

\theoremstyle{plain}
\newtheorem{thm}{\protect\theoremname}\newtheorem{cor}{\protect\corollaryname}\newtheorem{prop}{\protect\propositionname}\providecommand{\corollaryname}{Corollary}
\providecommand{\propositionname}{Proposition}
\providecommand{\theoremname}{Theorem}

\makeatother

\providecommand\theoremname{Theorem}
\providecommand\corollaryname{Corollary}

\providecommand\propositionname{Proposition}
\begin{document}
\title{GMM under finite-population asymptotics: instrumental variables and
regression adjustment}
\author{Haruo Kakehi}
\address{Department of Economics, University of Wisconsin-Madison, 1180 Observatory
Dr., Madison, 53706, WI, USA.}
\email{haruo.kakehi@wisc.edu}
\author{Yukitoshi Matsushita}
\address{Graduate School of Economics, Hitotsubashi University, 2-1 Naka, Kunitachi,
Tokyo 186-8601, Japan.}
\email{matsushita.y@r.hit-u.ac.jp}
\author{Taisuke Otsu}
\address{Department of Economics, London School of Economics, Houghton Street,
London, WC2A 2AE, UK.}
\email{t.otsu@lse.ac.uk}
\thanks{Matsushita acknowledges financial support from the JSPS KAKENHI (23K01331).}
\begin{abstract}
This paper extends the finite-population asymptotic framework of Abadie
\emph{et al.} (2014, 2020) to the generalized method of moments (GMM)
estimator for moment condition models. Motivating examples include
instrumental variable estimation and regression adjustment under randomized
controlled trials. We establish consistency and asymptotic normality
of the GMM estimator and propose asymptotically conservative variance
estimators. We also study GMM weighting, misspecification, and over-identification
testing under finite-population asymptotics. Notably, even though
the optimally weighted GMM estimator is infeasible under finite-population
asymptotics, the regression adjustment estimator in a randomized controlled
trial is shown to be asymptotically equivalent to the optimally weighted
GMM estimator. A simulation study and an empirical application illustrate
the usefulness of our GMM theory.
\end{abstract}

\maketitle

\section{Introduction}

Since the seminal works by Abadie \emph{et al.} (2014, 2020), there
has been growing interest in finite-population inference that accounts
for design-based uncertainty in empirical economic analysis. The design-based
perspective for investigating econometric or statistical methods has
been prevalent in randomized experiments (e.g., Neyman, 1923; Rosenbaum,
2002; Freedman, 2008a, 2008b). However, this literature typically
focuses on design-based uncertainty arising from random assignment
while treating the study population as fixed. On the other hand, an
extensive statistical literature exists on finite-population asymptotics,
taking into account sampling variation (see Prášková and Sen, 2009,
for an overview), although this body of work generally does not incorporate
design-based uncertainty. Abadie \emph{et al.} (2020) developed an
alternative inference framework by incorporating both design- and
sampling-based uncertainty. While their framework is restricted to
the case of linear regression, the recent literature considers design-based
uncertainty in different settings, including M-estimators (Xu, 2021),
spatial correlation (Xu and Wooldridge, 2022), and networks (Sakamoto
and Shimizu, 2025). Our paper is closely related to Xu (2021), which
extends Abadie \emph{et al.} (2020) to the M-estimation setting.

In this paper, we extend the existing results on finite-population
inference to situations where estimands of interest are defined by
over-identified moment conditions. Moment condition models are widely
used in empirical economic analysis, including instrumental variable
estimation and the analysis of randomized experiments. There is a
rich literature on the generalized method of moments (GMM) under conventional
infinite-population asymptotics (see, e.g., Hall, 2005, for an overview).
Therefore, it is of interest to develop GMM theory under finite-population
asymptotics that accounts for both sampling- and design-based uncertainty.

We study the asymptotic properties of the GMM estimator for moment
condition models under finite-population asymptotics. We derive the
consistency and asymptotic normality of the GMM estimator. In particular,
we find that its finite-population asymptotic variance takes a different
form from the conventional infinite-population asymptotic variance,
which is analogous to findings in just-identified models (e.g., Abadie
\emph{et al.}, 2020; Xu, 2021). Since our asymptotic variance for
the GMM estimator involves a component that is not consistently estimable,
we propose two asymptotically conservative variance estimators: one
is the conventional variance estimator, and the other is an adaptation
of the variance estimator in Abadie \emph{et al.} (2020) to the GMM
context. We also extend the finite-population GMM theory to globally
and locally misspecified models. Furthermore, we study the choice
of the GMM weight matrix under the finite-population asymptotic framework.
The inverse of the covariate-adjusted covariance estimator minimizes
the corresponding conservative asymptotic variance bound, although
the true asymptotic variances under this weight and the conventional
GMM weight cannot generally be ranked. To assess the choice between
these weights, we provide a sensitivity analysis based on an estimable
breakdown value. We also study an over-identification test and show
that the limiting distribution is generally a weighted chi-squared
distribution.

Our primary motivating examples are instrumental variable (IV) regression
models, which play a central role in the analysis of causal effects
in randomized experiments with noncompliance (e.g., Imbens and Angrist,
1994; Angrist, Imbens and Rubin, 1996). The recent design-based literature
has focused on inference in such experiments (e.g., Kang, Peck and
Keele, 2018; Jiang et al., 2024; Ren, 2024; Rambachan and Roth, 2025).
In this setting, our contribution is to extend the design-based framework
to allow for non-binary and multiple instruments and to incorporate
both sampling and design uncertainty in a unified GMM setting. Multiple-instrument
IV models also provide a natural setting in which over-identification
restrictions may be globally misspecified under treatment-effect heterogeneity
as in Lee (2018), even when the instruments themselves are valid.
In this case, we show that the GMM pseudo-true value depends on the
limiting weight matrix, and that estimation of the weight matrix can
contribute an additional first-order term to the asymptotic distribution.
We also provide a finite-population counterpart of the misspecification-robust
variance formula for a fixed weight.

Another important implication of the proposed GMM theory is that it
provides a framework to evaluate the efficiency gain from regression
adjustment for estimating the average treatment effect under a randomized
controlled trial. Regression adjustments for baseline covariates are
widely applied in practice and are also highly encouraged by regulatory
agencies to achieve more credible and efficient inference on causal
effects. Under the design-based setup, Lin (2013) suggested running
a regression of the observed outcome on the treatment variable, covariates,
and their interactions to guarantee an efficiency gain over the simple
difference-in-means estimator. Negi and Wooldridge (2021) established
an analogous efficiency gain result for regression adjustment under
the sampling-based or super-population setup. As emphasized in the
literature, such an efficiency gain for regression adjustment is guaranteed
regardless of correct specification of the linear model, which is
known as the no-harm property. The regression adjustment estimator
for the average treatment effect can be interpreted as a GMM estimator
for certain plug-in moment conditions, and an application of our GMM
theory shows that the same efficiency ordering continues to hold when
both sampling- and design-based uncertainty are present. Notably,
even though the optimally weighted GMM estimator is generally infeasible
under finite-population asymptotics, the regression adjustment estimator
for the average treatment effect in a randomized controlled trial
turns out to be asymptotically equivalent to the optimally weighted
GMM estimator.

The rest of the paper is organized as follows. Section \ref{sec:setup}
presents the basic setup and GMM estimator. In Section \ref{sub:RA-ex},
we discuss the regression adjustment estimator as an example. Section
\ref{sec:GMM} presents our main theoretical results on the asymptotic
properties of the GMM estimator under finite-population asymptotics.
Furthermore, Section \ref{sub:misspec} extends our finite-population
asymptotics to the situation where the moment conditions are misspecified,
Section \ref{sub:choice} discusses the choice of the GMM weight matrix
and presents a sensitivity diagnostic for the variance comparison,
and Section \ref{sub:overid} considers an over-identification test.
In Section \ref{sec:RA}, we apply the proposed GMM theory to investigate
the properties of the regression adjustment estimator in a randomized
controlled trial. Section \ref{sec:num} illustrates the proposed
GMM theory through a simulation study of an over-identified IV model
in Section \ref{sub:sim}. Finally, Section \ref{sub:app} provides
a real data application using the Opportunity Knocks experiment of
Angrist, Oreopoulos and Williams (2014). The proofs of our main results
are contained in Appendix \ref{app:A}.

\section{Setup and estimator }\label{sec:setup}

We first introduce our basic setup based on Abadie \emph{et al.} (2020)
and Xu (2021). Consider a finite population $\{X_{i},z_{i},Y_{i}\}^{M}_{i=1}$
of size $M$. For each unit $i=1,\ldots,M$, let $X_{i}$ be assignment
variables, $z_{i}$ be a vector of observed attributes, and $Y_{i}$
be a vector of outcome variables. For example, $X_{i}$ is a treatment
assignment in an experiment or an indicator for policy intervention,
and $z_{i}$ contains age, gender, and socioeconomic status of unit
$i$. Throughout the paper, we treat $z_{i}$ as fixed, and represent
outcomes as $Y_{i}=y_{i}(X_{i})$ using a potential outcome function
$y_{i}(\cdot)$, following the finite-population framework of Abadie
\emph{et al.} (2020). Formally, all variables depend on the population
size $M$, but we suppress this dependence for notational simplicity.
From the finite population $\{X_{i},z_{i},Y_{i}\}^{M}_{i=1}$, we
observe a random sample by Bernoulli sampling. Let $R_{i}$ be a Bernoulli
random variable, which equals one when unit $i$ is sampled, and zero
otherwise. Let $N=\sum^{M}_{i=1}R_{i}$ be the sample size, which
is random.\footnote{We note that in the finite-population asymptotic framework adopted
in this paper (as well as Abadie \emph{et al.,} 2020, and Xu, 2021),
the sample size $N$ and also the number of treated units $\sum^{M}_{i=1}R_{i}X_{i}$
are treated as random. On the other hand, in much of the statistics
literature on design-based analysis (e.g., Neyman, 1923, Lin, 2013,
and Li and Ding, 2017), the asymptotic analysis is typically conducted
conditionally on the number of treated units. Also note that all sample
averages and estimators below are defined on the event $\{N>0\}$.
Assumption (1) below guarantees $\mathbb{P}\{N=0\}=(1-\rho_{M})^{M}\to0$,
so the definitions on the event $\{N=0\}$ do not affect our asymptotic
results.}

We consider estimands defined as solutions to just- or over-identified
moment conditions. Let $g_{i}(X_{i},\theta):=g(X_{i},z_{i},y_{i}(X_{i}),\theta)$,
where $g$ is a $k$-dimensional vector of moment functions, $\theta$
is a $p$-dimensional parameter vector, and $k\ge p$. The estimand
of interest $\theta^{*}_{M}$ is defined as the unique solution to
\begin{equation}
\frac{1}{M}\sum^{M}_{i=1}\mathbb{E}[g_{i}(X_{i},\theta^{*}_{M})]=0.\label{eq:mom}
\end{equation}
The expectation $\mathbb{E}[\cdot]$ is taken with respect to the
assignment variables $X_{i}$. Existing finite-population asymptotic
results, including Abadie \emph{et al.} (2020) and Xu (2021), are
restricted to the just-identified case ($k=p$). Our framework extends
these studies by allowing both just- and over-identified moment conditions
($k\ge p$).

We focus primarily on the over-identified case and consider the GMM
estimator 
\begin{equation}
\hat{\theta}_{N}(W_{N})=\arg\min_{\theta\in\Theta}\left\{ \frac{1}{N}\sum^{M}_{i=1}R_{i}g_{i}(X_{i},\theta)\right\} ^{\prime}W_{N}\left\{ \frac{1}{N}\sum^{M}_{i=1}R_{i}g_{i}(X_{i},\theta)\right\} ,\label{eq:GMM}
\end{equation}
where $\Theta$ is the parameter space and $W_{N}$ is a $k\times k$
weight matrix. The choice of $W_{N}$ is discussed in Section \ref{sub:choice}.
We begin with two motivating examples before presenting the theoretical
properties of the estimator.

\begin{ex} {[}Instrumental variable{]} Kang, Peck and Keele (2018)
and Ren (2024) studied inference for LATEs in just-identified settings
under design-based frameworks. Our framework extends finite-population
asymptotics to over-identified settings with continuous instruments.
An important example of over-identified IV is studied by Lee (2018).
Under treatment effect heterogeneity, multiple instruments can identify
distinct instrument-specific LATEs. Consequently, the conventional
over-identified IV moment condition may be misspecified even when
each instrument is valid. Finite-population asymptotics also affect
the variance and weight matrix through a non-estimable component characterized
below, which has implications for statistical inference. \end{ex}

\begin{ex} {[}Auxiliary information{]} The second motivating example
is the use of auxiliary information on population moments, as in Imbens
and Lancaster (1994). This situation is relevant when the sampling
fraction is non-negligible and researchers can supplement the main
sample with administrative records, census aggregates, or large auxiliary
data sets. Examples include large-scale experiments where sample representativeness
is important (Muralidharan and Niehaus, 2017; Duflo and Banerjee,
2017) and the Integrated Public Use Microdata Series (IPUMS) data,
which contains a 10\% sample of the U.S. Census.

To formalize this idea, suppose that the moment function $g_{i}(X_{i},\theta,\eta_{M})$
depends on an auxiliary finite-population quantity $\eta_{M}$. If
$\eta_{M}$ is known from administrative records or population aggregates,
our framework incorporates auxiliary information. The same first-order
argument applies when $\eta_{M}$ is estimated from a large auxiliary
sample. Let $\hat{\eta}_{N_{a}}$ be an estimator based on an auxiliary
sample of size $N_{a}$. Assume $N/N_{a}\to0$. Under standard regularity
conditions, the estimation error in $\hat{\eta}_{N_{a}}$ is asymptotically
negligible, so replacing $\eta_{M}$ by $\hat{\eta}_{N_{a}}$ has
no first-order effect on the asymptotic distribution.\footnote{Under standard regularity conditions, we have $\sqrt{N}(\hat{\eta}_{N_{a}}-\eta_{M})=\sqrt{\frac{N}{N_{a}}}\sqrt{N_{a}}(\hat{\eta}_{N_{a}}-\eta_{M})=o_{p}(1)$,
which implies that 
\[
\frac{1}{\sqrt{N}}\sum^{M}_{i=1}R_{i}g_{i}(X_{i},\theta,\hat{\eta}_{N_{a}})=\frac{1}{\sqrt{N}}\sum^{M}_{i=1}R_{i}g_{i}(X_{i},\theta,\eta_{M})+o_{p}(1).
\]
}\end{ex}

\subsection{Regression adjustment }\label{sub:RA-ex}

An important application of our GMM theory under finite-population
asymptotics is regression adjustment to estimate the average treatment
effect under a randomized controlled trial. In this subsection, let
$X_{i}\in\{0,1\}$ be the assignment indicator with common assignment
probability $\pi=\mathbb{P}\{X_{i}=1\}\in(0,1)$ ($X_{i}=1$ corresponds
to the treatment, and $X_{i}=0$ corresponds to the control), and
$y_{i}(x)$ be a potential outcome for the treatment $x\in\{0,1\}$.
For each unit $i$ with $R_{i}=1$, we observe the randomly assigned
treatment $X_{i}$, the observable outcome $Y_{i}=y_{i}(X_{i})$,
and a vector of non-random pre-treatment covariates $z_{i}$ (without
a constant term).

In this setup, we wish to conduct inference on the average treatment
effect $\theta^{*}_{M}=\frac{1}{M}\sum^{M}_{i=1}\{y_{i}(1)-y_{i}(0)\}$
using the observable data $\{X_{i},Y_{i},z_{i}\}_{i:R_{i}=1}$. Let
\begin{eqnarray*}
\bar{Y}_{1} & = & \frac{1}{N_{1}}\sum^{M}_{i=1}R_{i}X_{i}Y_{i},\qquad\bar{Y}_{0}=\frac{1}{N_{0}}\sum^{M}_{i=1}R_{i}(1-X_{i})Y_{i},\\
\bar{z}_{1} & = & \frac{1}{N_{1}}\sum^{M}_{i=1}R_{i}X_{i}z_{i},\qquad\bar{z}_{0}=\frac{1}{N_{0}}\sum^{M}_{i=1}R_{i}(1-X_{i})z_{i},\qquad\bar{z}=\frac{1}{N}\sum^{M}_{i=1}R_{i}z_{i},
\end{eqnarray*}
where $N_{1}=\sum^{M}_{i=1}R_{i}X_{i}$ and $N_{0}=\sum^{M}_{i=1}R_{i}(1-X_{i})$
are the numbers of treated and control units in the sample, respectively.
The estimators in this subsection are well defined on the event $\{N_{1}>0,N_{0}>0\}$,
whose probability converges to one. Popular estimators of $\theta^{*}_{M}$
are
\begin{eqnarray}
\hat{\theta}_{\mathrm{DIM}} & = & \bar{Y}_{1}-\bar{Y}_{0},\qquad\text{(difference-in-means)}\nonumber \\
\hat{\theta}_{\mathrm{RA}} & = & \bar{Y}_{1}-\bar{Y}_{0}-(\hat{\beta}^{\prime}_{1},-\hat{\beta}^{\prime}_{0})\left(\begin{array}{c}
\bar{z}_{1}-\bar{z}\\
\bar{z}_{0}-\bar{z}
\end{array}\right),\qquad\text{(regression adjustment)}\label{eq:RA}
\end{eqnarray}
where 
\begin{align*}
\hat{\beta}_{1} & =\left(\sum^{M}_{i=1}R_{i}X_{i}(z_{i}-\bar{z}_{1})(z_{i}-\bar{z}_{1})^{\prime}\right)^{-1}\sum^{M}_{i=1}R_{i}X_{i}(z_{i}-\bar{z}_{1})Y_{i},\\
\hat{\beta}_{0} & =\left(\sum^{M}_{i=1}R_{i}(1-X_{i})(z_{i}-\bar{z}_{0})(z_{i}-\bar{z}_{0})^{\prime}\right)^{-1}\sum^{M}_{i=1}R_{i}(1-X_{i})(z_{i}-\bar{z}_{0})Y_{i}.
\end{align*}

Under the design-based setup, Lin (2013) showed that the regression
adjustment estimator $\hat{\theta}_{\mathrm{RA}}$ is asymptotically
more efficient than the difference-in-means estimator $\hat{\theta}_{\mathrm{DIM}}$
without assuming correct specification of the linear regression model.
Under the sampling-based (or super-population) setup, Negi and Wooldridge
(2021) established an analogous efficiency guarantee for regression
adjustment. We investigate whether this efficiency improvement for
regression adjustment continues to hold under the present finite-population
asymptotic framework.

The GMM theory in this paper provides a unified framework to study
asymptotic properties of $\hat{\theta}_{\mathrm{DIM}}$ and $\hat{\theta}_{\mathrm{RA}}$
under finite-population asymptotics. In particular, define the population
moment function for the assignment probability $\pi$ by 
\[
g_{i}(X_{i},\theta;\pi)=\left(\begin{array}{c}
\dfrac{X_{i}Y_{i}}{\pi}-\dfrac{(1-X_{i})Y_{i}}{1-\pi}-\theta\\
\left(\dfrac{X_{i}}{\pi}-\dfrac{1-X_{i}}{1-\pi}\right)z_{i}
\end{array}\right).
\]
The corresponding feasible plug-in moment function is $\hat{g}_{i}(X_{i},\theta)=g_{i}(X_{i},\theta;\hat{\pi})$,
where $\hat{\pi}=N_{1}/N$. The population moment condition is therefore
given by (\ref{eq:mom}), whereas estimation uses $\hat{\pi}$ in
place of $\pi$. Then the difference-in-means estimator $\hat{\theta}_{\mathrm{DIM}}$
and an asymptotically equivalent counterpart of the regression adjustment
estimator $\hat{\theta}_{\mathrm{RA}}$ can be interpreted as the
GMM estimators with suitably chosen weight matrices. In Section \ref{sec:RA}
below, we apply the GMM theory under finite-population asymptotics
to show that $\hat{\theta}_{\mathrm{RA}}$ is asymptotically more
efficient than $\hat{\theta}_{\mathrm{DIM}}$ and is indeed optimal
within the class of GMM estimators.

\section{GMM under finite-population asymptotics }\label{sec:GMM}

We now study the large-sample properties of the GMM estimator $\hat{\theta}_{N}(W_{N})$
in our asymptotic framework. We impose the following assumptions.

\begin{asm}$\quad$ 
\begin{enumerate}
\item $\{X_{i}\}^{M}_{i=1}$ is independent but not necessarily identically
distributed. $\{R_{i}\}^{M}_{i=1}$ is an independent and identically
distributed sequence of Bernoulli random variables with $\rho_{M}=\mathbb{P}\{R_{i}=1\}\in(0,1]$
for every $M$ and satisfying $M\rho_{M}\to\infty$ and $\rho_{M}\to\rho\in[0,1]$
as $M\to\infty$. Furthermore, $\{X_{i}\}^{M}_{i=1}$ and $\{R_{i}\}^{M}_{i=1}$
are independent. 
\item $\Theta$ is compact. $W_{N}$ converges in probability to a positive
definite matrix $W$.\\
 $\lim_{M\to\infty}\frac{1}{M}\sum^{M}_{i=1}\mathbb{E}[g_{i}(X_{i},\theta)]=0$
is uniquely satisfied at $\theta=\theta^{*}:=\lim_{M\to\infty}\theta^{*}_{M}$.
$g_{i}(x,\theta)$ is continuous at each $\theta\in\Theta$ for almost
every $x$, and $\sup_{i,M}\mathbb{E}[\sup_{\theta\in\Theta}||g_{i}(X_{i},\theta)||^{4}]<\infty$.
There exist functions $h_{1}(\cdot)$ and $b_{1i}(\cdot)$ such that
$\lim_{u\to0}h_{1}(u)=0$, $\sup_{i,M}\mathbb{E}[b_{1i}(X_{i})]<\infty$,
and $||g_{i}(X_{i},\theta)-g_{i}(X_{i},\theta_{1})||\le b_{1i}(X_{i})h_{1}(||\theta-\theta_{1}||)$
for each $\theta,\theta_{1}\in\Theta$. 
\item $\theta^{*}\in\mathrm{int}(\Theta)$. $g_{i}(x,\theta)$ is continuously
differentiable on $\mathrm{int}(\Theta)$ for almost every $x$, and
$G:=\lim_{M\to\infty}\frac{1}{M}\sum^{M}_{i=1}\mathbb{E}\left[\frac{\partial g_{i}(X_{i},\theta^{*}_{M})}{\partial\theta^{\prime}}\right]$
has full column rank. The following limits exist and are finite:
\[
\Omega=\lim_{M\to\infty}\frac{1}{M}\sum^{M}_{i=1}\mathbb{E}[g_{i}(X_{i},\theta^{*}_{M})g_{i}(X_{i},\theta^{*}_{M})^{\prime}],\qquad\Delta=\lim_{M\to\infty}\frac{1}{M}\sum^{M}_{i=1}\mathbb{E}[g_{i}(X_{i},\theta^{*}_{M})]\mathbb{E}[g_{i}(X_{i},\theta^{*}_{M})]^{\prime}.
\]
Furthermore, $\Omega-\rho\Delta$ is positive definite. There exist
functions $h_{2}(\cdot)$ and $b_{2i}(\cdot)$ and a neighborhood
$\mathcal{N}$ around $\theta^{*}$ such that\\
 $\sup_{i,M}\mathbb{E}[\sup_{\theta\in\mathcal{N}}||\partial g_{i}(X_{i},\theta)/\partial\theta^{\prime}||^{2}]<\infty$,
$\lim_{u\to0}h_{2}(u)=0$, $\sup_{i,M}\mathbb{E}[b_{2i}(X_{i})]<\infty$,
and $||\partial g_{i}(X_{i},\theta)/\partial\theta^{\prime}-\partial g_{i}(X_{i},\theta_{1})/\partial\theta^{\prime}||\le b_{2i}(X_{i})h_{2}(||\theta-\theta_{1}||)$
for each $\theta,\theta_{1}\in\mathcal{N}$. 
\end{enumerate}
\end{asm}

Assumption (1) is on the sampling framework, which is also employed
by Abadie \emph{et al.} (2020) and Xu (2021). This implies the sample
size $N=\sum^{M}_{i=1}R_{i}$ is random and its expectation $\mathbb{E}[N]=M\rho_{M}$
diverges. The case $\rho=0$ nests the conventional negligible-sampling-fraction
asymptotics, whereas $\rho>0$ retains the finite-population correction.
Assumptions (2) and (3) collect regularity conditions on the weight
matrix $W_{N}$ and the moment function $g_{i}$. These are natural
adaptations of the conventional GMM theory to our finite-population
setup. Assumption (2) is used to derive the consistency of the GMM
estimator, and Assumption (3) contains additional conditions to establish
asymptotic normality.

Under the above assumptions, the asymptotic properties of the GMM
estimator $\hat{\theta}_{N}(W_{N})$ are obtained as follows. 
\begin{thm}
\label{thm:GMM}$\quad$ 
\begin{enumerate}
\item Under Assumptions (1)-(2), it holds that $\hat{\theta}_{N}(W_{N})-\theta^{*}_{M}\overset{p}{\to}0$. 
\item Under Assumptions (1)-(3), it holds that 
\begin{equation}
\sqrt{N}(\hat{\theta}_{N}(W_{N})-\theta^{*}_{M})\overset{d}{\to}N(0,V_{\mathrm{GMM}}(W)),\label{eq:an}
\end{equation}
where 
\[
V_{\mathrm{GMM}}(W)=(G^{\prime}WG)^{-1}G^{\prime}W(\Omega-\rho\Delta)WG(G^{\prime}WG)^{-1}.
\]
\end{enumerate}
\end{thm}
Theorem \ref{thm:GMM} (1) says that the GMM estimator $\hat{\theta}_{N}(W_{N})$
is consistent for the population parameter $\text{\ensuremath{\theta^{*}_{M}}}$,
and Theorem \ref{thm:GMM} (2) derives its asymptotic distribution.
Compared with the conventional infinite-population asymptotics, the
main difference is the presence of the additional term ``$\rho\Delta$''
in the asymptotic variance. Letting $V_{\mathrm{C}}(W)=(G^{\prime}WG)^{-1}G^{\prime}W\Omega WG(G^{\prime}WG)^{-1}$
be the asymptotic variance of the GMM estimator under the conventional
infinite-population asymptotics, and $V_{\mathrm{A}}(W)=\rho(G^{\prime}WG)^{-1}G^{\prime}W\Delta WG(G^{\prime}WG)^{-1}$
be the additional component, the finite-population asymptotic variance
can be written as $V_{\mathrm{GMM}}(W)=V_{\mathrm{C}}(W)-V_{\mathrm{A}}(W)$.
Since $\Delta$ is positive semi-definite, $V_{\mathrm{A}}(W)$ is
also positive semi-definite and $V_{\mathrm{GMM}}(W)$ is always no
larger than the conventional variance $V_{\mathrm{C}}(W)$ in the
matrix sense (denoted by $V_{\mathrm{GMM}}(W)\le_{\mathrm{pd}}V_{\mathrm{C}}(W)$).
Although $V_{\mathrm{C}}(W)$ can be consistently estimated (as shown
below), the component $\Delta$ and thus the variance $V_{\mathrm{GMM}}(W)$
cannot be consistently estimated in general.

As in Abadie \emph{et al.} (2020) and Xu (2021), we propose conservative
estimators for $V_{\mathrm{GMM}}(W)$. The first variance estimator
is a consistent estimator of the conventional variance $V_{\mathrm{C}}(W)$
given by 
\[
\hat{V}_{\mathrm{C}}(W_{N})=(\hat{G}^{\prime}W_{N}\hat{G})^{-1}\hat{G}^{\prime}W_{N}\hat{\Omega}W_{N}\hat{G}(\hat{G}^{\prime}W_{N}\hat{G})^{-1},
\]
where $\hat{G}=\frac{1}{N}\sum^{M}_{i=1}R_{i}\partial g_{i}(X_{i},\hat{\theta}_{N}(W_{N}))/\partial\theta^{\prime}$
and $\hat{\Omega}=\frac{1}{N}\sum^{M}_{i=1}R_{i}g_{i}(X_{i},\hat{\theta}_{N}(W_{N}))g_{i}(X_{i},\hat{\theta}_{N}(W_{N}))^{\prime}$.
The second variance estimator is based on an estimator of a lower
bound of $\Delta$, given by
\[
\hat{\Delta}_{Z}=\frac{1}{N}\sum^{M}_{i=1}R_{i}\hat{P}^{\prime}z_{i}z^{\prime}_{i}\hat{P},
\]
where $\hat{P}=\left(\sum^{M}_{i=1}R_{i}z_{i}z^{\prime}_{i}\right)^{-1}\left(\sum^{M}_{i=1}R_{i}z_{i}g_{i}(X_{i},\hat{\theta}_{N}(W_{N}))^{\prime}\right)$
and $\widehat{\rho}=N/M$. Since $\widehat{\rho}-\rho_{M}=o_{p}(1)$,
replacing the sampling probability by the observed sampling fraction
does not change any first-order result. Then the second variance estimator
is written as 
\[
\hat{V}_{Z}(W_{N})=(\hat{G}^{\prime}W_{N}\hat{G})^{-1}\hat{G}^{\prime}W_{N}(\hat{\Omega}-\hat{\rho}\hat{\Delta}_{Z})W_{N}\hat{G}(\hat{G}^{\prime}W_{N}\hat{G})^{-1}.
\]
The asymptotic properties of these variance estimators are presented
as follows. 
\begin{thm}
\label{thm:var}$\quad$ 
\begin{description}
\item [{(1)}] Under Assumptions (1)-(3), $\hat{V}_{\mathrm{C}}(W_{N})\overset{p}{\to}V_{\mathrm{C}}(W)$. 
\item [{(2)}] In addition to Assumptions (1)-(3), assume that $\lim_{M\to\infty}\frac{1}{M}\sum^{M}_{i=1}\mathbb{E}[g_{i}(X_{i},\theta^{*}_{M})]z^{\prime}_{i}$
exists, $\lim_{M\to\infty}\frac{1}{M}\sum^{M}_{i=1}z_{i}z^{\prime}_{i}$
exists and is nonsingular, and $\sup_{i,M}||z_{i}||<\infty$. Then
$\hat{\Delta}_{Z}$ converges in probability to a positive semi-definite
matrix $\Delta_{Z}$ such that $\Delta_{Z}\le_{\mathrm{pd}}\Delta$. 
\item [{(3)}] Under the assumptions of Part (2), $V_{\mathrm{GMM}}(W)\le_{\mathrm{pd}}V_{Z}(W)\le_{\mathrm{pd}}V_{\mathrm{C}}(W)$,
where $V_{Z}(W)=(G^{\prime}WG)^{-1}G^{\prime}W(\Omega-\rho\Delta_{Z})WG(G^{\prime}WG)^{-1}$.
Furthermore, $V_{Z}((\Omega-\rho\Delta_{Z})^{-1})\le_{\mathrm{pd}}V_{Z}(\Omega^{-1})$. 
\end{description}
\end{thm}
Since the proofs of the remaining parts are similar to those of Xu
(2021, Theorems 2.2 and 3.1), they are omitted.\footnote{The uniform boundedness condition on $z_{i}$ in Theorem 2 (2) is
imposed only to keep the proof parallel to Xu (2021, Theorem 3.1).
This condition can be relaxed to $\lim_{M\to\infty}\frac{1}{M}\sum^{M}_{i=1}||z_{i}||^{4+\delta}<\infty$
as in Abadie \emph{et al.} (2020, Assumption 5).} Theorem \ref{thm:var} (1) says that the conventional variance estimator
$\hat{V}_{\mathrm{C}}(W_{N})$ is still consistent for the variance
component $V_{\mathrm{C}}(W)$ under finite-population asymptotics.
Theorem \ref{thm:var} (2) establishes that $\hat{\Delta}_{Z}$ consistently
estimates a lower bound for $\Delta$, and Theorem \ref{thm:var}
(3) establishes the conservativeness of the resulting variance estimator
$\hat{V}_{Z}(W_{N})$. Theorem \ref{thm:var} (3) also clarifies the
relationships among the limiting variance matrices. Although we cannot
consistently estimate the asymptotic variance $V_{\mathrm{GMM}}(W)$
of the GMM estimator, we can provide asymptotically conservative estimators
$\hat{V}_{\mathrm{C}}(W_{N})$ and $\hat{V}_{Z}(W_{N})$. We recommend
using the second estimator $\hat{V}_{Z}(W_{N})$ under the finite-population
asymptotic framework because it is less conservative than the first
estimator $\hat{V}_{\mathrm{C}}(W_{N})$.

\subsection{Misspecified moment conditions }\label{sub:misspec}

When the moment conditions are over-identified $(k>p)$, there may
be no value of $\theta$ that sets all population moments to zero.
A leading example is an IV model with multiple valid instruments and
heterogeneous treatment effects. Different instruments can identify
different instrument-specific LATEs, so a common over-identified IV
restriction can be globally misspecified even though instrument validity
is not in question (Lee, 2018). In that case, the GMM estimand is
a weight-dependent approximation to the instrument-specific LATEs
rather than a common LATE. This subsection extends the preceding finite-population
GMM framework to allow for misspecified moment conditions.

To analyze misspecified moment conditions, we impose the following
assumptions. Let $\bar{g}_{M}(\theta)=M^{-1}\sum^{M}_{i=1}\mathbb{E}[g_{i}(X_{i},\theta)]$
and $\bar{G}_{M}(\theta)=M^{-1}\sum^{M}_{i=1}\mathbb{E}[D_{i}(X_{i},\theta)]$,
where $D_{i}(X_{i},\theta)=\partial g_{i}(X_{i},\theta)/\partial\theta^{\prime}$.
For a limiting weight matrix $W$, define 
\[
Q_{M}(\theta,W)=\bar{g}_{M}(\theta)^{\prime}W\bar{g}_{M}(\theta),\qquad\theta^{\#}_{M}(W)=\arg\min_{\theta\in\Theta}Q_{M}(\theta,W).
\]

\begin{asm}$\quad$ 
\begin{description}
\item [{(2)'}] Assumption (2), except for correct specification and identification
by (\ref{eq:mom}), holds. For each sufficiently large $M$, $\theta^{\#}_{M}(W)$
is the unique minimizer of $Q_{M}(\theta,W)$, and $\theta^{\#}_{M}(W)\to\theta^{\#}(W)\in\mathrm{int}(\Theta)$. 
\item [{(3)'}] The function $g_{i}(x,\theta)$ is twice continuously differentiable
in a neighborhood $\mathcal{N}$ of $\theta^{\#}(W)$ for almost every
$x$. Let $D^{(2)}_{i}(X_{i},\theta)=\partial\mathrm{vec}D_{i}(X_{i},\theta)/\partial\theta^{\prime}$.
The first and second derivatives satisfy the same uniform square-integrability
and Lipschitz-envelope conditions as in Assumption (3). In particular,
for $j=2,3$ there are envelopes $b_{ji}$ and moduli $h_{j}$ such
that $\sup_{i,M}\mathbb{E}[b_{ji}(X_{i})]<\infty$, $h_{j}(u)\to0$,
and 
\[
\begin{aligned}\|D_{i}(X_{i},\theta)-D_{i}(X_{i},\theta_{1})\| & \le b_{2i}(X_{i})h_{2}(\|\theta-\theta_{1}\|),\\
\|D^{(2)}_{i}(X_{i},\theta)-D^{(2)}_{i}(X_{i},\theta_{1})\| & \le b_{3i}(X_{i})h_{3}(\|\theta-\theta_{1}\|),
\end{aligned}
\]
uniformly for $\theta,\theta_{1}\in\mathcal{N}$. The limits defined
below exist and are finite, and $\mathcal{H}^{\#}$ is nonsingular. 
\end{description}
\end{asm}

To simplify the presentation, let 
\[
\begin{aligned}\mu^{\#}_{M} & =\bar{g}_{M}(\theta^{\#}_{M}),\qquad G^{\#}_{M}=\bar{G}_{M}(\theta^{\#}_{M}),\qquad G^{(2)\#}_{M}=\frac{1}{M}\sum^{M}_{i=1}\mathbb{E}[D^{(2)}_{i}(X_{i},\theta^{\#}_{M})],\\
\mathcal{H}^{\#}_{M} & =G^{\#\prime}_{M}WG^{\#}_{M}+(I_{p}\otimes\mu^{\#\prime}_{M}W)G^{(2)\#}_{M},
\end{aligned}
\]
where the dependence of $\theta^{\#}_{M}$ on $W$ is suppressed.
Their limits are denoted by $\mu^{\#}$, $G^{\#}$, $G^{(2)\#}$,
and $\mathcal{H}^{\#}$. The second term in $\mathcal{H}^{\#}_{M}$
is absent under correct specification and is the source of the familiar
misspecified-GMM Hessian in Hall and Inoue (2003). The finite-population
variance can be stated transparently by collecting the first-order
score contribution of unit $i$: 
\[
\psi^{\#}_{iM}=G^{\#\prime}_{M}W\{g_{i}(X_{i},\theta^{\#}_{M})-\mu^{\#}_{M}\}+\{D_{i}(X_{i},\theta^{\#}_{M})-G^{\#}_{M}\}^{\prime}W\mu^{\#}_{M}.
\]
Note that $M^{-1}\sum_{i}\mathbb{E}[\psi^{\#}_{iM}]=0$ by construction.
We also define 
\[
\Omega^{\#}_{\psi}=\lim_{M\to\infty}\frac{1}{M}\sum^{M}_{i=1}\mathbb{E}[\psi^{\#}_{iM}\psi^{\#\prime}_{iM}],\qquad\Delta^{\#}_{\psi}=\lim_{M\to\infty}\frac{1}{M}\sum^{M}_{i=1}\mathbb{E}[\psi^{\#}_{iM}]\mathbb{E}[\psi^{\#}_{iM}]^{\prime}.
\]

Under these assumptions, we obtain the following extension of Theorem
\ref{thm:GMM}. 
\begin{cor}
\label{cor:mis}$\quad$ 
\begin{enumerate}
\item Under Assumptions (1) and (2)', $\hat{\theta}_{N}(W_{N})-\theta^{\#}_{M}(W)\overset{p}{\to}0$. 
\item Under Assumptions (1), (2)', and (3)', if $\sqrt{N}(W_{N}-W)=o_{p}(1)$
(in particular, for a fixed weight $W$), then 
\[
\sqrt{N}\{\hat{\theta}_{N}(W_{N})-\theta^{\#}_{M}(W)\}\overset{d}{\to}N(0,V_{\mathrm{mis}}(W)),
\]
where $V_{\mathrm{mis}}(W)=(\mathcal{H}^{\#})^{-1}(\Omega^{\#}_{\psi}-\rho\Delta^{\#}_{\psi})(\mathcal{H}^{\#})^{-1\prime}$. 
\item More generally, suppose $W_{N}$ is estimated and 
\[
\left(\frac{1}{\sqrt{N}}\sum^{M}_{i=1}R_{i}\psi^{\#}_{iM},\sqrt{N}(W_{N}-W)\right)\overset{d}{\to}(\mathcal{Z}_{\psi},\mathcal{Z}_{W})
\]
jointly, where $Var(\mathcal{Z}_{\psi})=\Omega^{\#}_{\psi}-\rho\Delta^{\#}_{\psi}$
and $\mathcal{Z}_{W}$ is a random $k\times k$ matrix. Then 
\[
\sqrt{N}\{\hat{\theta}_{N}(W_{N})-\theta^{\#}_{M}(W)\}\overset{d}{\to}-(\mathcal{H}^{\#})^{-1}\{\mathcal{Z}_{\psi}+G^{\#\prime}\mathcal{Z}_{W}\mu^{\#}\}.
\]
\end{enumerate}
\end{cor}
Corollary \ref{cor:mis} makes two differences from the correctly
specified case explicit. First, global misspecification makes the
pseudo-true value depend on the limiting weight matrix. Second, estimation
of the weight matrix is first-order relevant through $G^{\#\prime}\mathcal{Z}_{W}\mu^{\#}$.
For a fixed weight, the sampling ratio enters the variance through
the explicit correction $-\rho\Delta^{\#}_{\psi}$. Thus, the conventional
misspecified GMM sandwich formula does not directly apply under finite-population
asymptotics.

A conservative variance estimator is obtained from the sample analogue
of $V_{\mathrm{mis}}(W)$. For a fixed weight, let $\hat{\mu}=\hat{g}_{N}(\hat{\theta}_{N})$,
$\hat{G}=\hat{G}_{N}(\widehat{\theta}_{N})$, $\hat{\mathcal{H}}$
be the sample analogue of $\mathcal{H}^{\#}$, and 
\[
\hat{\psi}_{i}=\hat{G}^{\prime}W\{g_{i}(X_{i},\hat{\theta}_{N})-\hat{\mu}\}+\{D_{i}(X_{i},\widehat{\theta}_{N})-\hat{G}\}^{\prime}W\hat{\mu},\qquad\hat{\Omega}_{\psi}=\frac{1}{N}\sum^{M}_{i=1}R_{i}\hat{\psi}_{i}\hat{\psi}^{\prime}_{i}.
\]
It follows that $\hat{\mathcal{H}}^{-1}\hat{\Omega}_{\psi}\hat{\mathcal{H}}^{-1\prime}$
is asymptotically conservative. As in Theorem \ref{thm:var}, a less
conservative estimator is obtained by estimating a lower bound of
$\Delta^{\#}_{\psi}$. Let 
\[
\hat{P}_{\psi}=\left(\sum^{M}_{i=1}R_{i}z_{i}z^{\prime}_{i}\right)^{-1}\left(\sum^{M}_{i=1}R_{i}z_{i}\hat{\psi}^{\prime}_{i}\right),\qquad\hat{\Delta}_{\psi,Z}=\frac{1}{N}\sum^{M}_{i=1}R_{i}\hat{P}^{\prime}_{\psi}z_{i}z^{\prime}_{i}\hat{P}_{\psi}.
\]
Under the same conditions as in Theorem \ref{thm:var}, the probability
limit $\Delta_{\psi,Z}$ of $\hat{\Delta}_{\psi,Z}$ satisfies $0\le_{\mathrm{pd}}\Delta_{\psi,Z}\le_{\mathrm{pd}}\Delta^{\#}_{\psi}$.
Hence the less conservative estimator replaces $\hat{\Omega}_{\psi}$
by $\hat{\Omega}_{\psi}-\hat{\rho}\hat{\Delta}_{\psi,Z}$, where $\hat{\rho}=N/M$.
\begin{rem}
{[}Linear IV case; connection to Lee (2018){]} A useful special case
is the multiple IV setting studied by Lee (2018), where valid instruments
may identify different instrument-specific LATEs under heterogeneous
treatment effects, leading to globally misspecified moment conditions.
In the linear IV case, we have $G^{(2)\#}=0$ so that $\mathcal{H}^{\#}=G^{\#\prime}WG^{\#}$.
Corollary \ref{cor:mis} therefore provides the asymptotic distribution
of the over-identified IV estimator under global misspecification.
In particular, the contribution from the random Jacobian term contained
in $\psi^{\#}_{iM}$ is first-order relevant under global misspecification.
The variance estimator $\hat{\mathcal{H}}^{-1}\hat{\Omega}_{\psi}\hat{\mathcal{H}}^{-1\prime}$
described above provides a finite-population counterpart of the misspecification-robust
variance estimator in Lee (2018).
\end{rem}
\begin{rem}
{[}Local misspecification{]} Global misspecification changes the probability
limit and, for estimated weights, the first-order distribution. The
effects are different under a sequence of local misspecification.
Suppose there is a deterministic sequence $\theta^{0}_{M}$ and a
fixed vector $\delta$ such that 
\[
\bar{g}_{M}(\theta^{0}_{M})=(M\rho_{M})^{-1/2}\delta+o((M\rho_{M})^{-1/2}),
\]
while the Jacobian converges to $G$. Since $N/(M\rho_{M})\overset{p}{\to}1$,
the GMM estimator with a fixed weight satisfies 
\[
\sqrt{N}\{\hat{\theta}_{N}(W)-\theta^{0}_{M}\}\overset{d}{\to}N(-A(W)\delta,V_{\mathrm{GMM}}(W)),\qquad A(W)=(G^{\prime}WG)^{-1}G^{\prime}W.
\]
Equivalently, $\theta^{\#}_{M}(W)-\theta^{0}_{M}=-(M\rho_{M})^{-1/2}A(W)\delta+o((M\rho_{M})^{-1/2})$,
and centering at the local pseudo-true value removes the mean shift.
Thus local misspecification leaves the asymptotic variance unchanged
and affects only the limiting mean; this contrasts with the weight-dependent
estimand and additional first-order terms under global misspecification. 
\end{rem}

\subsection{Choice of GMM weight }\label{sub:choice}

The choice of the weight matrix is a feature specific to over-identified
GMM and therefore the main practical distinction from the just-identified
finite-population results in Abadie \emph{et al.} (2020) and Xu (2021).
The asymptotic analysis above treats the probability limit of $W_{N}$
as given. Since the asymptotic variance $V_{\mathrm{GMM}}(W)$ differs
from the one under conventional infinite-population asymptotics, it
is interesting to investigate the choice of the weight matrix under
the current setup.

The variance $V_{\mathrm{GMM}}(W)$ is minimized by $W_{\mathrm{opt}}=(\Omega-\rho\Delta)^{-1}$
in the matrix sense. However, due to the component $\Delta$, a consistent
estimator of $W_{\mathrm{opt}}$ is not available in general. Motivated
by the discussion following \ref{thm:var}, we can consider two feasible
weights, $\hat{\Omega}^{-1}$ and $(\hat{\Omega}-\hat{\rho}\hat{\Delta}_{Z})^{-1}$.
Theorems \ref{thm:GMM} and \ref{thm:var} imply 
\begin{eqnarray*}
\sqrt{N}(\hat{\theta}_{N}(\hat{\Omega}^{-1})-\theta^{*}_{M}) & \overset{d}{\to} & N(0,V_{\mathrm{GMM}}(\Omega^{-1})),\\
\sqrt{N}(\hat{\theta}_{N}((\hat{\Omega}-\hat{\rho}\hat{\Delta}_{Z})^{-1})-\theta^{*}_{M}) & \overset{d}{\to} & N(0,V_{\mathrm{GMM}}((\Omega-\rho\Delta_{Z})^{-1})).
\end{eqnarray*}
We note that neither of these asymptotic variances dominates the other
in general.\footnote{To illustrate, consider the case of $k=2$, $p=1$, and $\rho=0.8$
with 
\[
\Omega=\left[\begin{array}{cc}
5 & 0.5\\
0.5 & 5
\end{array}\right],\qquad\Delta=\left[\begin{array}{cc}
4 & 0.1\\
0.1 & 4.8
\end{array}\right],\qquad G=\left[\begin{array}{c}
1\\
1
\end{array}\right],
\]
which yields $V_{\mathrm{GMM}}(\Omega^{-1})\approx0.95$. On the other
hand, $V_{\mathrm{GMM}}((\Omega-\rho\Delta_{Z})^{-1})$ varies with
$\Delta_{Z}$. For example, let $\Delta_{Z}=\left[\begin{array}{cc}
c & 0.1\\
0.1 & 2
\end{array}\right]$ for some $c>0$ satisfying $0\leq\Delta_{Z}\le_{\mathrm{pd}}\Delta$.
When $c=1$, it holds $V_{\mathrm{GMM}}((\Omega-\rho\Delta_{Z})^{-1})\approx0.92<V_{\mathrm{GMM}}(\Omega^{-1})$.
However, when $c=3$, it holds $V_{\mathrm{GMM}}((\Omega-\rho\Delta_{Z})^{-1})\approx1.01>V_{\mathrm{GMM}}(\Omega^{-1})$.} However, Theorem \ref{thm:var} (3) implies 
\begin{equation}
V_{\mathrm{GMM}}((\Omega-\rho\Delta_{Z})^{-1})\le_{\mathrm{pd}}V_{Z}((\Omega-\rho\Delta_{Z})^{-1})\le_{\mathrm{pd}}V_{Z}(\Omega^{-1})\le_{\mathrm{pd}}V_{\mathrm{C}}(\Omega^{-1}).\label{eq:var}
\end{equation}
Based on these relationships, for inference on $\theta^{*}_{M}$,
we recommend using the point estimator $\hat{\theta}_{N}((\hat{\Omega}-\hat{\rho}\hat{\Delta}_{Z})^{-1})$
together with the asymptotic variance estimator $\hat{V}_{Z}((\hat{\Omega}-\hat{\rho}\hat{\Delta}_{Z})^{-1})$.
Although this choice does not necessarily yield a smaller asymptotic
variance for the point estimator, it yields a less conservative variance
estimator.
\begin{rem}
{[}Power property{]} To illustrate the power property of inference
based on $\hat{\theta}_{N}((\hat{\Omega}-\hat{\rho}\hat{\Delta}_{Z})^{-1})$
and $\hat{V}_{Z}((\hat{\Omega}-\hat{\rho}\hat{\Delta}_{Z})^{-1})$,
let us consider the case where $\theta^{*}_{M}$ is scalar, and compare
the $t$-statistics: 
\[
T_{\rho}=\frac{\hat{\theta}_{N}((\hat{\Omega}-\hat{\rho}\hat{\Delta}_{Z})^{-1})-c}{\sqrt{\hat{V}_{Z}((\hat{\Omega}-\hat{\rho}\hat{\Delta}_{Z})^{-1})/N}},\qquad T_{0}=\frac{\hat{\theta}_{N}(\hat{\Omega}^{-1})-c}{\sqrt{\hat{V}_{Z}(\hat{\Omega}^{-1})/N}},
\]
for testing the null hypothesis $H_{0}:\theta^{*}_{M}=c$ against
the one-sided alternative $H_{1}:\theta^{*}_{M}>c$. Under the local
alternative hypothesis $H_{1M}:\theta^{*}_{M}=c+(M\rho_{M})^{-1/2}h$
for some $h>0$, an adaptation of the proofs of Theorems \ref{thm:GMM}
and \ref{thm:var} implies 
\[
T_{\rho}\overset{d}{\to}N(\mu_{\rho},V_{\rho}),\qquad T_{0}\overset{d}{\to}N(\mu_{0},V_{0}),
\]
where $\mu_{\rho}=\frac{h}{\sqrt{V_{Z}((\Omega-\rho\Delta_{Z})^{-1})}}$,
$\mu_{0}=\frac{h}{\sqrt{V_{Z}(\Omega^{-1})}}$, $V_{\rho}=\frac{V_{\mathrm{GMM}}((\Omega-\rho\Delta_{Z})^{-1})}{V_{Z}((\Omega-\rho\Delta_{Z})^{-1})}$,
and $V_{0}=\frac{V_{\mathrm{GMM}}(\Omega^{-1})}{V_{Z}(\Omega^{-1})}$.
Thus, if we employ the same critical value $\xi$, the limiting local
power functions of the tests based on these statistics are written
as 
\[
\mathbb{P}\{T_{\rho}\ge\xi\}\to1-\Phi\left(\frac{\xi-\mu_{\rho}}{\sqrt{V_{\rho}}}\right),\qquad\mathbb{P}\{T_{0}\ge\xi\}\to1-\Phi\left(\frac{\xi-\mu_{0}}{\sqrt{V_{0}}}\right),
\]
where $\Phi(\cdot)$ is the cumulative distribution function of the
$N(0,1)$ distribution. Although the noncentrality parameters satisfy
$\mu_{\rho}\ge\mu_{0}$, we cannot rank the variance terms $V_{\rho}$
and $V_{0}$. Therefore, the local powers of the two tests cannot
be ranked in general. The test based on $T_{\rho}$ has no lower local
power than the one based on $T_{0}$ whenever $\frac{\xi-\mu_{\rho}}{\sqrt{V_{\rho}}}\le\frac{\xi-\mu_{0}}{\sqrt{V_{0}}}$. 
\end{rem}
\begin{rem}
{[}Diagnostics for weight choice{]} We provide a diagnostic for assessing
the range of uncertainty about the non-estimable component of $\Delta$
over which an efficiency gain from the adjusted weight $(\hat{\Omega}-\hat{\rho}\hat{\Delta_{Z}})^{-1}$
can be guaranteed. Consider a scalar contrast $\ell^{\prime}\theta^{*}_{M}$,
and let $a_{0}=\Omega^{-1}G(G^{\prime}\Omega^{-1}G)^{-1}\ell$ and
$a_{Z}=(\Omega-\rho\Delta_{Z})^{-1}G(G^{\prime}(\Omega-\rho\Delta_{Z})^{-1}G)^{-1}\ell$.
For $s\in[0,1]$, define
\[
\mathcal{D}(s)=\{\tilde{\Delta}:\Delta_{Z}\leq_{\mathrm{pd}}\tilde{\Delta}\leq_{\mathrm{pd}}\Delta_{Z}+s(\Omega-\Delta_{Z})\}.
\]
When $s=0$, $\mathcal{D}(s)$ reduces to the singleton $\{\Delta_{Z}\}$,
whereas when $s=1$, it recovers the baseline restriction $\Delta_{Z}\leq_{\mathrm{pd}}\tilde{\Delta}\leq_{\mathrm{pd}}\Omega$.
Let $d=(a_{0}-a_{Z})^{\prime}(\Omega-\rho\Delta_{Z})(a_{0}-a_{Z})$,
and $\lambda^{+}$ be the sum of the positive eigenvalues of $(\Omega-\Delta_{Z})^{1/2}(a_{0}a^{\prime}_{0}-a_{Z}a^{\prime}_{Z})(\Omega-\Delta_{Z})^{1/2}$.
Then it follows that
\[
\inf_{\tilde{\Delta}\in\mathcal{D}(s)}\ell^{\prime}\{V_{\mathrm{GMM}}(\Omega^{-1};\tilde{\Delta})-V_{\mathrm{GMM}}((\Omega-\rho\Delta_{Z})^{-1};\tilde{\Delta})\}\ell=d-\rho s\lambda^{+}.
\]
Therefore, when $d,\lambda^{+}>0$, define the breakdown value
\[
s^{*}=\min\left\{ 1,\frac{d}{\rho\lambda^{+}}\right\} .
\]
For every $0\leq s<s^{*}$ and $\tilde{\Delta}\in\mathcal{D}(s)$,
the adjusted weight has a smaller asymptotic variance, i.e.,
\[
\ell^{\prime}V_{\mathrm{GMM}}((\Omega-\rho\Delta_{Z})^{-1};\tilde{\Delta})\ell<\ell^{\prime}V_{\mathrm{GMM}}(\Omega^{-1};\tilde{\Delta})\ell.
\]
If $s^{*}<1$, the sensitivity restriction alone no longer guarantees
a strictly smaller asymptotic variance for $s\geq s^{*}$. If $s^{*}=1$,
the variance improvement is guaranteed for all $s<1$. A larger value
of $s^{*}$ therefore indicates that the variance reduction from using
the adjusted weight can be guaranteed over a wider range of uncertainty
about the non-estimable component $\Delta-\Delta_{Z}$.
\end{rem}

\subsection{Over-identification test }\label{sub:overid}

Similar to parameter hypothesis testing, the asymptotic behavior of
the over-identification test is also different from the conventional
super-population setting. In this subsection, we consider the following
over-identifying restriction test statistic 
\[
J_{N}(W_{N})=N\left(\frac{1}{N}\sum^{M}_{i=1}R_{i}g_{i}(X_{i},\hat{\theta}_{N}(W_{N}))\right)^{\prime}W_{N}\left(\frac{1}{N}\sum^{M}_{i=1}R_{i}g_{i}(X_{i},\hat{\theta}_{N}(W_{N}))\right).
\]
The asymptotic property of $J_{N}(W_{N})$ is obtained as follows. 
\begin{cor}
\label{cor:overid} Suppose Assumptions (1)-(3) hold true. 
\begin{enumerate}
\item If $W_{N}\overset{p}{\to}W=(\Omega-\rho\Delta)^{-1}$ as $M\to\infty$,
then $J_{N}(W_{N})\overset{d}{\to}\chi^{2}_{k-p}$. 
\item If $W_{N}\overset{p}{\to}W\neq(\Omega-\rho\Delta)^{-1}$ as $M\to\infty$,
then $J_{N}(W_{N})\overset{d}{\to}\sum^{k-p}_{j=1}\lambda_{j}\chi^{2}_{1,j}$,
where $\{\chi^{2}_{1,j}\}^{k-p}_{j=1}$ are mutually independent $\chi^{2}_{1}$
random variables and $\{\lambda_{j}\}^{k-p}_{j=1}$ are the non-zero
eigenvalues of $(\Omega-\rho\Delta)\{W-WG(G^{\prime}WG)^{-1}G^{\prime}W\}$. 
\end{enumerate}
\end{cor}
Corollary \ref{cor:overid} shows that the limiting distribution of
the statistic $J_{N}(W_{N})$ is in general characterized by a weighted
sum of chi-squared random variables. Due to the component $\Delta$,
we cannot consistently estimate the eigenvalues $\{\lambda_{j}\}^{k-p}_{j=1}$.
A practical approach is to substitute $\Delta_{Z}$ for $\Delta$
and work with the eigenvalues of $(\Omega-\rho\Delta_{Z})\{W-WG(G^{\prime}WG)^{-1}G^{\prime}W\}$,
denoted by $\{\lambda_{Z,j}\}^{k-p}_{j=1}$. By Theorem \ref{thm:var}
and the Courant-Fischer-Weyl min-max principle, we have $\lambda_{Z,j}\geq\lambda_{j}$
for all $j=1,\ldots,k-p$. Thus, we can employ the estimated quantiles
of $\sum^{k-p}_{j=1}\lambda_{Z,j}\chi^{2}_{1,j}$ as conservative
critical values for testing the over-identifying restrictions (i.e.,
eq. (\ref{eq:mom}) holds true at some $\theta^{*}_{M}\in\Theta$).

\section{Regression adjustment }\label{sec:RA}

In this section, we apply the GMM theory presented in the last section
to compare the difference-in-means estimator $\hat{\theta}_{\mathrm{DIM}}$
and the regression adjustment estimator $\hat{\theta}_{\mathrm{RA}}$
for estimating the average treatment effect under random assignment
of the binary treatment. Recall the setup and notation in Section
\ref{sub:RA-ex}, where the population moment is $g_{i}(X_{i},\theta;\pi)$,
and its feasible plug-in counterpart is 
\begin{equation}
\hat{g}_{i}(X_{i},\theta)=\left(\begin{array}{c}
\dfrac{X_{i}Y_{i}}{\hat{\pi}}-\dfrac{(1-X_{i})Y_{i}}{1-\hat{\pi}}-\theta\\
\left(\dfrac{X_{i}}{\hat{\pi}}-\dfrac{1-X_{i}}{1-\widehat{\pi}}\right)z_{i}
\end{array}\right),\label{eq:g-RA}
\end{equation}
with $\hat{\pi}=N_{1}/N$. Using this moment function, we consider
the GMM estimator for $\theta^{*}_{M}$: 
\[
\hat{\theta}_{N}(W)=\arg\min_{\theta}\left\{ \frac{1}{N}\sum^{M}_{i=1}R_{i}\hat{g}_{i}(X_{i},\theta)\right\} ^{\prime}W\left\{ \frac{1}{N}\sum^{M}_{i=1}R_{i}\hat{g}_{i}(X_{i},\theta)\right\} ,
\]
where $W=\left(\begin{array}{cc}
W_{1} & W^{\prime}_{12}\\
W_{12} & W_{2}
\end{array}\right)$ with scalar $W_{1}$. Let $\bar{z}_{M}=\frac{1}{M}\sum^{M}_{i=1}z_{i}$,
$\theta^{*}_{Mk}=\frac{1}{M}\sum^{M}_{i=1}y_{i}(k)$, and $v_{i}(k)=y_{i}(k)-\theta^{*}_{Mk}$
for $k=0,1$. Then mild regularity conditions guarantee 
\begin{equation}
\frac{1}{\sqrt{N}}\sum^{M}_{i=1}R_{i}\hat{g}_{i}(X_{i},\theta^{*}_{M})=\frac{1}{\sqrt{N}}\sum^{M}_{i=1}R_{i}\left(\begin{array}{c}
\frac{X_{i}v_{i}(1)}{\pi}-\frac{(1-X_{i})v_{i}(0)}{1-\pi}\\
\frac{(X_{i}-\pi)(z_{i}-\bar{z}_{M})}{\pi(1-\pi)}
\end{array}\right)+o_{p}(1),\label{eq:ghat}
\end{equation}
by using $\hat{\pi}=\frac{\sum^{M}_{i=1}R_{i}X_{i}}{\sum^{M}_{i=1}R_{i}}$,
$\frac{1}{N_{1}}\sum^{M}_{i=1}R_{i}X_{i}Y_{i}=\theta^{*}_{M1}+o_{p}(1)$,
and $\frac{1}{N_{0}}\sum^{M}_{i=1}R_{i}(1-X_{i})Y_{i}=\theta^{*}_{M0}+o_{p}(1)$
(due to the law of large numbers). Combining this with $G=\lim_{M\to\infty}\frac{1}{M}\sum^{M}_{i=1}\mathbb{E}\left[\frac{\partial g_{i}(X_{i},\theta^{*}_{M};\pi)}{\partial\theta^{\prime}}\right]=(-1,0^{\prime})^{\prime}$
for the population counterpart of (\ref{eq:g-RA}), an argument analogous
to that for Theorem \ref{thm:GMM} implies 
\begin{eqnarray*}
\mathrm{asy.var}[\sqrt{N}(\hat{\theta}_{N}(W)-\theta^{*}_{M})] & = & (G^{\prime}WG)^{-1}G^{\prime}W\Omega^{*}WG(G^{\prime}WG)^{-1}\\
 & = & (1,W^{-1}_{1}W^{\prime}_{12})\left(\begin{array}{cc}
\Omega^{*}_{1} & \Omega^{*\prime}_{12}\\
\Omega^{*}_{12} & \Omega^{*}_{2}
\end{array}\right)\left(\begin{array}{c}
1\\
W^{-1}_{1}W_{12}
\end{array}\right),
\end{eqnarray*}
where 
\begin{align*}
\Omega^{*}_{1} & =\lim_{M\to\infty}\frac{1}{M}\sum^{M}_{i=1}\left\{ \frac{v_{i}(1)^{2}}{\pi}+\frac{v_{i}(0)^{2}}{1-\pi}\right\} -\rho\lim_{M\to\infty}\frac{1}{M}\sum^{M}_{i=1}(v_{i}(1)-v_{i}(0))^{2}\\
\Omega^{*}_{2} & =\lim_{M\to\infty}\frac{1}{M}\sum^{M}_{i=1}\frac{(z_{i}-\bar{z}_{M})(z_{i}-\bar{z}_{M})^{\prime}}{\pi(1-\pi)},\\
\Omega^{*}_{12} & =\lim_{M\to\infty}\frac{1}{M}\sum^{M}_{i=1}\left\{ \frac{(z_{i}-\bar{z}_{M})v_{i}(1)}{\pi}+\frac{(z_{i}-\bar{z}_{M})v_{i}(0)}{1-\pi}\right\} .
\end{align*}
Although we cannot consistently estimate the optimal weight $\Omega^{*-1}=\left(\begin{array}{cc}
\Omega^{*}_{1} & \Omega^{*\prime}_{12}\\
\Omega^{*}_{12} & \Omega^{*}_{2}
\end{array}\right)^{-1}$ (especially $\Omega^{*}_{1}$), an inspection of the above asymptotic
variance reveals that $\hat{\theta}_{N}(W)$ can achieve the same
asymptotic efficiency as the (infeasible) optimally weighted GMM estimator
$\hat{\theta}_{N}(\Omega^{*-1})$ when the weight $W$ satisfies 
\begin{equation}
W^{-1}_{1}W^{\prime}_{12}=-\Omega^{*\prime}_{12}\Omega^{*-1}_{2}.\label{eq:key}
\end{equation}

A key observation is that condition (\ref{eq:key}) does not involve
$\Omega^{*}_{1}$, which is typically difficult to estimate consistently
from the observed data under the present setup.\footnote{This feature follows directly from the asymptotic variance formula
of the GMM. To see this, letting $c=W^{-1}_{1}W_{12}$, we have 
\begin{align*}
\mathrm{asy.var}[\sqrt{N}(\hat{\theta}_{N}(W)-\theta^{*}_{M})] & =(1,c^{\prime})\begin{pmatrix}\Omega^{*}_{1} & \Omega^{*\prime}_{12}\\
\Omega^{*}_{12} & \Omega^{*}_{2}
\end{pmatrix}\begin{pmatrix}1\\
c
\end{pmatrix}=\Omega^{*}_{1}+2c'\Omega^{*}_{12}+c'\Omega^{*}_{2}c.
\end{align*}
Minimizing this expression with respect to $c$ gives $c=-\Omega^{*-1}_{2}\Omega^{*}_{12}$,
or equivalently $W^{-1}_{1}W^{\prime}_{12}=-\Omega^{*\prime}_{12}\Omega^{*-1}_{2}$,
which is precisely the condition in (\ref{eq:key}).} Indeed, we can show that the probability limit of the weight 
\begin{equation}
W_{\mathrm{RA}}=\left(\begin{array}{cc}
1 & \frac{1}{N}\sum^{M}_{i=1}R_{i}\left(\frac{X_{i}}{\hat{\pi}}-\frac{1-X_{i}}{1-\hat{\pi}}\right)^{2}Y_{i}(z_{i}-\bar{z})^{\prime}\\
\frac{1}{N}\sum^{M}_{i=1}R_{i}\left(\frac{X_{i}}{\hat{\pi}}-\frac{1-X_{i}}{1-\hat{\pi}}\right)^{2}Y_{i}(z_{i}-\bar{z}) & \frac{1}{N}\sum^{M}_{i=1}R_{i}\left(\frac{X_{i}}{\hat{\pi}}-\frac{1-X_{i}}{1-\hat{\pi}}\right)^{2}(z_{i}-\bar{z})(z_{i}-\bar{z})^{\prime}
\end{array}\right)^{-1},\label{eq:WRA}
\end{equation}
satisfies the condition in (\ref{eq:key}) so that the resulting GMM
estimator $\hat{\theta}_{N}(W_{\mathrm{RA}})$ has the same asymptotic
efficiency as the optimally weighted GMM estimator $\hat{\theta}_{N}(\Omega^{*-1})$.
It can also be shown that $\hat{\theta}_{N}(W_{\mathrm{RA}})$ is
asymptotically equivalent to the regression adjustment estimator $\hat{\theta}_{\mathrm{RA}}$
in (\ref{eq:RA}).

Moreover, since the difference-in-means estimator $\hat{\theta}_{\mathrm{DIM}}$
corresponds to $\hat{\theta}_{N}(W_{\mathrm{DIM}})$ with $W_{\mathrm{DIM}}=\left(\begin{array}{cc}
1 & {\bf 0}\\
{\bf 0} & {\bf 0}
\end{array}\right)$, we can conclude that the difference-in-means estimator is no more
efficient than $\hat{\theta}_{\mathrm{RA}}$ for all $\rho\in[0,1]$
under finite-population asymptotics. These results are summarized
in the following proposition. Recall $V_{\mathrm{GMM}}(W)=(G^{\prime}WG)^{-1}G^{\prime}W\Omega^{*}WG(G^{\prime}WG)^{-1}$. 
\begin{prop}
Consider the setup of Section \ref{sub:RA-ex} and this section. Suppose
(\ref{eq:ghat}) holds and $\tilde{g}_{i}(X_{i},\theta^{*}_{M})=\left(\begin{array}{c}
\frac{X_{i}v_{i}(1)}{\pi}-\frac{(1-X_{i})v_{i}(0)}{1-\pi}\\
\frac{(X_{i}-\pi)(z_{i}-\bar{z}_{M})}{\pi(1-\pi)}
\end{array}\right)$ satisfies Assumptions (1)-(3). Then 
\end{prop}
\begin{enumerate}
\item $\sqrt{N}(\hat{\theta}_{N}(W_{\mathrm{RA}})-\theta^{*}_{M})\overset{d}{\to}N(0,V_{\mathrm{GMM}}(\Omega^{*-1}))$
and $\sqrt{N}(\hat{\theta}_{N}(W_{\mathrm{RA}})-\hat{\theta}_{\mathrm{RA}})=o_{p}(1)$. 
\item $\sqrt{N}(\hat{\theta}_{N}(W_{\mathrm{DIM}})-\theta^{*}_{M})\overset{d}{\to}N(0,V_{\mathrm{GMM}}(W_{\mathrm{DIM}}))$
and $V_{\mathrm{GMM}}(W_{\mathrm{DIM}})\ge V_{\mathrm{GMM}}(\Omega^{*-1})$. 
\end{enumerate}
The proof follows by adapting the argument used in Theorem \ref{thm:GMM}
to $\tilde{g}_{i}(X_{i},\theta^{*}_{M})$ combined with (\ref{eq:ghat}).
The proof of the statement $\sqrt{N}(\hat{\theta}_{N}(W_{\mathrm{RA}})-\hat{\theta}_{\mathrm{RA}})=o_{p}(1)$
follows by applying the argument of Tsiatis \emph{et al.} (2008) to
the finite-population asymptotic setup. In contrast to the general
weight-selection problem discussed in Section \ref{sub:choice}, the
optimality condition in (\ref{eq:key}) does not depend on $\Omega^{*}_{1}$,
which is not consistently estimable under finite-population asymptotics.
This proposition shows that the regression adjustment estimator $\hat{\theta}_{N}(W_{\mathrm{RA}})$
is no less efficient than the difference-in-means estimator $\hat{\theta}_{N}(W_{\mathrm{DIM}})$
for every $\rho\in[0,1]$. This result extends the efficiency comparison
of Lin (2013) and Negi and Wooldridge (2021) to the present finite-population
asymptotic framework. The result relies on the particular moment structure
in the regression adjustment problem and does not generally apply
to the choice of the GMM weight matrix considered in Section \ref{sub:choice}.

Note that even though the point estimator $\hat{\theta}_{N}(W_{\mathrm{RA}})$
or $\hat{\theta}_{\mathrm{RA}}$ achieves asymptotic optimality in
the class of GMM estimators, its asymptotic variance $V_{\mathrm{GMM}}(\Omega^{*-1})=(G^{\prime}\Omega^{*-1}G)^{-1}$
involves $\Omega^{*}$, which cannot be consistently estimated in
general. Thus, the resulting inference methods are conservative as
in the general case.

Finally, the above results can be extended to stratified randomized
experiments (e.g., Wang, Wang and Liu, 2023) with a finite number
of strata, where for each group $g=1,\ldots,G$, the treatment assignment
mechanism satisfies $\mathbb{P}\{X_{i}=1|G_{i}=g\}=\pi_{g}$ with
the group indicator $G_{i}$. In this case, let $N_{g}=\sum^{M}_{i=1}R_{i}\mathbb{I}\{G_{i}=g\}$,
$N_{1g}=\sum^{M}_{i=1}R_{i}\mathbb{I}\{G_{i}=g\}X_{i}$, and $\hat{\pi}_{g}=N_{1g}/N_{g}$.
The feasible plug-in moment is
\[
\hat{g}_{i}(X_{i},\theta)=\left(\begin{array}{c}
{\displaystyle \sum^{G}_{g=1}\mathbb{I}\{G_{i}=g\}\left(\frac{X_{i}Y_{i}}{\hat{\pi}_{g}}-\frac{(1-X_{i})Y_{i}}{1-\hat{\pi}_{g}}\right)-\theta}\\[4pt]
{\displaystyle \left\{ \mathbb{I}\{G_{i}=g\}\left(\frac{X_{i}}{\hat{\pi}_{g}}-\frac{1-X_{i}}{1-\hat{\pi}_{g}}\right)z_{i}\right\} ^{G}_{g=1}}
\end{array}\right).
\]
Note that the factor $N_{g}/N$ does not appear in the unit-level
first moment because it is generated when $N^{-1}\sum_{i}R_{i}\hat{g}_{i}$
is averaged over observations. The corresponding population moment
$g_{i}(X_{i},\theta;\pi)$ replaces each $\hat{\pi}_{g}$ by $\pi_{g}$.
The same argument as above then establishes an analogous efficiency
result for stratified regression adjustment.

\section{Numerical illustration }\label{sec:num}

\subsection{Simulation }\label{sub:sim}

This subsection illustrates our theoretical results through Monte
Carlo simulation of an over-identified IV model. We model the endogenous
treatment and outcome as
\begin{eqnarray*}
D_{i} & = & X^{\prime}_{i}\pi+v_{i},\\
Y_{i} & = & (1+1.25w_{i}+\alpha q_{i})D_{i}+u_{i},
\end{eqnarray*}

where $X_{i}=(X_{i1},\ldots,X_{iK})^{\prime}\in\mathbb{R}^{K}$ is
the instrument vector. The scalar $w_{i}$ is an observed covariate,
and $q_{i}$ captures unobserved heterogeneity of treatment effects.
For each population size $M\in\{500,1000,2000\}$, we set $w_{i}\in\{-1.5,0,1.5\}$
with finite-population probabilities 0.25, 0.50, and 0.25, respectively,
fixing the corresponding population proportions. We consider $K\in\{2,5\}$
and set the first-stage coefficient vector as $\pi=(6,1.05)^{\prime}$
for $K=2$ and $\pi=(9,1.10,0.25,0.25,0.25)^{\prime}$ for $K=5$.
The coefficient $\alpha$ in the outcome equation is set to 1 when
$w_{i}=0$; it remains 0.10 when $w_{i}=\pm1.5$. 

We generate the disturbances and unobserved heterogeneity jointly
as
\[
\begin{pmatrix}u_{i}\\
v_{i}\\
q_{i}
\end{pmatrix}\sim N\left(0,\begin{pmatrix}0.50 & 0.25 & 0\\
0.25 & 0.50 & 0\\
0 & 0 & 1
\end{pmatrix}\right),
\]
and normalize within each $w_{i}$ block to match these finite-population
moments. Correlation between $u_{i}$ and $v_{i}$ causes endogeneity.
Unit characteristics remain fixed across replications. Instrument
strength varies with $w_{i}$. In each replication, we independently
draw $e_{ij}\in\{-1,1\}$, each with probability 0.50, across units
and instruments and rescale them to obtain $X_{ij}$. For $K=2$,
the conditional variances of the first and second instruments are
1.75 and 0.25 when $w_{i}=\pm1.5$, and 0.25 and 3.25 when $w_{i}=0$,
respectively. For $K=5$, the conditional variance of the first instrument
is 1 in every block, while that of each remaining instrument is 0.25
when $w_{i}=\pm1.5$ and 5 when $w_{i}=0$. Independently of instrument
assignment, Bernoulli sampling uses $R_{i}\sim\mathrm{Bernoulli}(\rho)$
with $\rho\in\{0.25,0.50,0.75\}$.

We estimate $\theta=(\beta,\gamma)^{\prime}$ using the moment function
\[
g_{i}(X_{i},\theta)=\begin{pmatrix}X_{i}\\
w_{i}
\end{pmatrix}(Y_{i}-\beta D_{i}-\gamma w_{i}),
\]
with finite-population target $\theta^{*}_{M}=(\beta^{*}_{M},\gamma^{*}_{M})^{\prime}$.
We compare two-step GMM estimates of $\beta$ using weights $\hat{\Omega}^{-1}$
and $(\hat{\Omega}-\rho\hat{\Delta}_{Z})^{-1}$, estimating $\hat{\Delta}_{Z}$
using $(1,w_{i})^{\prime}$. We report standard errors and coverage
rates of the 95\% confidence intervals based on $V_{C}$ and $V_{Z}$,
the estimated breakdown value $s^{*}$, and rejection rates of conventional
and weighted over-identification tests at the 5\% level. Each design
uses $5,000$ Monte Carlo replications.

\begin{sidewaystable}[ph]
\vspace{10cm}

\caption{Simulation results ($K=2$)}

\begin{tabular}{c|ccc|ccc|ccc}
\hline 
 & \multicolumn{3}{c|}{$M=500$} & \multicolumn{3}{c|}{$M=1000$} & \multicolumn{3}{c}{$M=2000$}\tabularnewline
 & $\rho=0.25$  & $\rho=0.5$  & $\rho=0.75$  & $\rho=0.25$  & $\rho=0.5$  & $\rho=0.75$  & $\rho=0.25$  & $\rho=0.5$  & $\rho=0.75$\tabularnewline
\hline 
$\lvert\bar{\hat{\beta}}(\hat{\Omega}^{-1})-\beta^{*}_{M}\rvert$  & 0.0005 & 0.0001 & 0.0003 & 0.0007 & 0.0006 & 0.0001 & 0.0002 & 0.0004 & 0.0007\tabularnewline
$\lvert\bar{\hat{\beta}}((\hat{\Omega}-\rho\hat{\Delta}_{Z})^{-1})-\beta^{*}_{M}\rvert$  & 0.0010 & 0.0001 & 0.0004 & 0.0004 & 0.0009 & 0.0005 & 0.0001 & 0.0001 & 0.0002\tabularnewline
\hline 
SD $\hat{\beta}(\hat{\Omega}^{-1})$  & 0.1722 & 0.1054 & 0.0713 & 0.1217 & 0.0758 & 0.0512 & 0.0877 & 0.0527 & 0.0354\tabularnewline
SD $\hat{\beta}((\hat{\Omega}-\rho\hat{\Delta}_{Z})^{-1})$  & 0.1714 & 0.1026 & 0.0631 & 0.1204 & 0.0726 & 0.0435 & 0.0872 & 0.0510 & 0.0307\tabularnewline
$\sqrt{V_{\mathrm{GMM}}(\Omega^{-1})/N}$  & 0.1711 & 0.1055 & 0.0713 & 0.1211 & 0.0746 & 0.0504 & 0.0856 & 0.0528 & 0.0356\tabularnewline
$\sqrt{V_{\mathrm{GMM}}((\Omega-\rho\Delta_{Z})^{-1})/N}$  & 0.1701 & 0.1018 & 0.0617 & 0.1204 & 0.0720 & 0.0436 & 0.0851 & 0.0509 & 0.0308\tabularnewline
\hline 
SE $V_{\mathrm{C}}(\Omega^{-1})$  & 0.1897 & 0.1346 & 0.1100 & 0.1345 & 0.0952 & 0.0778 & 0.0952 & 0.0673 & 0.0550\tabularnewline
SE $V_{Z}(\Omega^{-1})$  & 0.1722 & 0.1078 & 0.0745 & 0.1219 & 0.0761 & 0.0526 & 0.0861 & 0.0538 & 0.0371\tabularnewline
SE $V_{Z}((\Omega-\rho\Delta_{Z})^{-1})$  & 0.1711 & 0.1035 & 0.0636 & 0.1210 & 0.0730 & 0.0446 & 0.0855 & 0.0515 & 0.0314\tabularnewline
\hline 
Coverage $V_{\mathrm{C}}(\Omega^{-1})$  & 0.9668 & 0.9850 & 0.9966 & 0.9686 & 0.9854 & 0.9960 & 0.9636 & 0.9862 & 0.9978\tabularnewline
Coverage $V_{Z}(\Omega^{-1})$  & 0.9474 & 0.9562 & 0.9576 & 0.9512 & 0.9502 & 0.9548 & 0.9434 & 0.9512 & 0.9600\tabularnewline
Coverage $V_{Z}((\Omega-\rho\Delta_{Z})^{-1})$  & 0.9466 & 0.9522 & 0.9500 & 0.9510 & 0.9462 & 0.9552 & 0.9426 & 0.9502 & 0.9524\tabularnewline
\hline 
$s^{*}$ & 0.5098 & 0.5935 & 0.7069 & 0.5255 & 0.6041 & 0.7175 & 0.5334 & 0.6095 & 0.7232\tabularnewline
\hline 
Chi-Squared & 0.0408 & 0.0326 & 0.0188 & 0.0382 & 0.0306 & 0.0250 & 0.0378 & 0.0318 & 0.0212\tabularnewline
Weighted Chi-Squared & 0.0494 & 0.0450 & 0.0384 & 0.0474 & 0.0456 & 0.0410 & 0.0442 & 0.0434 & 0.0390\tabularnewline
\end{tabular}
\end{sidewaystable}

Table 1 reports the absolute biases and standard deviations of the
point estimators, average standard errors, and empirical coverages
of the corresponding 95\% confidence intervals for $K=2$. We also
report the theoretical standard deviations $\sqrt{V_{\mathrm{GMM}}(\Omega^{-1})/N}$
and $\sqrt{V_{\mathrm{GMM}}((\Omega-\rho\Delta_{Z})^{-1})/N}$, calculated
using knowledge of the data generating process, together with the
weight-choice diagnostic and rejection rates of the over-identification
tests. The theoretical standard deviations use $N=M\rho$, whereas
estimated standard errors and confidence intervals use the realized
sample size. Table 3 in Appendix B reports the results for $K=5$,
which are qualitatively similar to those for $K=2$.

First, we compare the point estimators. Both $\hat{\beta}(\widehat{\Omega}^{-1})$
and $\widehat{\beta}((\widehat{\Omega}-\rho\widehat{\Delta}_{Z})^{-1})$
exhibit small absolute biases in all designs. Changing the weight
reduces the Monte Carlo standard deviation by 0.44--15.00\% relative
to the conventional weight. The reduction increases with $\rho$,
ranging from 0.44--1.02\% at $\rho=0.25$ to 11.48--15.00\% at $\rho=0.75$.
The Monte Carlo standard deviations are close to their theoretical
counterparts, supporting the accuracy of the asymptotic approximation.

Next, we compare the standard errors. Holding the conventional weight
fixed, replacing the conventional variance estimator with the design-based
variance estimator reduces average standard errors by 9.25--32.49\%.
Changing the weight further to $(\hat{\Omega}-\rho\hat{\Delta}_{Z})^{-1}$
yields an additional reduction of 0.62--15.32\%. Both reductions
become larger as $\rho$ increases. Thus, the design-based variance
correction and the weight change provide distinct improvements in
precision.

The conventional confidence intervals exhibit over-coverage, with
empirical coverage ranging from 0.9636 to 0.9978. Using the design-based
variance estimator with the conventional weight brings coverage closer
to 0.95, with values between 0.9434 and 0.9600. Combining the design-based
variance estimator with the adjusted weight yields coverage between
0.9426 and 0.9552. These intervals therefore maintain coverage close
to the nominal level while substantially reducing standard errors.

The Monte Carlo means of the estimated breakdown values $s^{*}$ range
from 0.5098 to 0.7232 and increase with $\rho$. Recall that $s^{*}$
is the largest sensitivity level below which variance improvement
is guaranteed for every $\widetilde{\Delta}\in\mathcal{D}(s)$, with
$s<s^{*}$. The larger estimated breakdown values provide stronger
support for using the adjusted weight. Thus, it suggests that the
variance gain holds over a wider range of uncertainty about $\Delta$
for larger $\rho$.

Finally, the conventional chi-squared over-identification test is
conservative, with null rejection rates between 0.0188 and 0.0408.
Its conservativeness becomes more pronounced as $\rho$ increases.
Using weighted chi-squared critical values brings rejection rates
closer to the nominal 5\% level in every design, with rates between
0.0384 and 0.0494. Some conservativeness remains at the larger sampling
fractions.

\subsection{Application }\label{sub:app}

We revisit the Opportunity Knocks (OK) experiment of Angrist, Oreopoulos
and Williams (2014), conducted at a Canadian commuter college. The
experiment tested whether grade-contingent cash awards and peer advising
improved students\textquoteright{} academic performance. In 2008,
2,129 first- and second-year students who had applied for financial
aid were invited to participate. Among the invitees, 1,271 completed
the baseline survey and consented to random assignment; 400 were assigned
to receive the offer. Treated students were offered \$100 for each
course grade of at least 70, plus \$20 for each percentage point above
70, as well as email-based peer advising. Assignment was stratified
by sex, year of study, and high-school GPA quartile, yielding 16 strata.
A key outcome is full-year GPA, which measures overall academic performance.
Table 6 of Angrist, Oreopoulos and Williams (2014) focuses on students
who correctly answered a five-course award-calculation question on
the pre-assignment survey, a subgroup intended to capture students
who understood the incentive scheme. None of the nine reported GPA
effects is statistically significant at conventional levels.

We reanalyze these nine full-year GPA comparisons. The original paper
estimates the additive OLS specification $Y_{i}=\alpha_{g(i)}+\tau X_{i}+z^{\prime}_{i}\beta+\varepsilon_{i}$,
where $g(i)$ indexes the randomization strata, $X_{i}$ indicates
assignment to the offer, and $z_{i}$ includes high-school grades,
an indicator for English as a first language, parental-education indicators,
and pre-assignment responses about the award rules. We reproduce the
original point estimates and standard errors, then apply the stratified
GMM regression adjustment of Section \ref{sec:RA} using the same
covariates. We report the point estimate $\hat{\theta}(W_{\mathrm{RA}})$
and compare $\hat{V}_{\mathrm{C}}(W_{\mathrm{RA}})$ with $\hat{V}_{Z}(W_{\mathrm{RA}})$,
holding the point estimate fixed. For the pooled GPA comparison, 803
students have complete outcome and covariate records, giving the sampling
fraction $\hat{\rho}$=803/846, and the subgroup columns use analogous
ratios. We take the 846 pre-assignment correct responders as the finite
population because they define the target subgroup in the original
Table 6 analysis.

\begin{sidewaystable}[ph]
\vspace{10cm}

\caption{Application results}

\begin{tabular}{c|ccc|ccc|ccc}
\hline 
 & \multicolumn{3}{c|}{Women} & \multicolumn{3}{c|}{Men} & \multicolumn{3}{c}{All}\tabularnewline
 & First Years & Second Years & All & First Years & Second Years & All & First Years & Second Years & All\tabularnewline
\hline 
Original  & \begin{cellvarwidth}[t]
\centering
-0.1069 

(0.0882)
\end{cellvarwidth} & \begin{cellvarwidth}[t]
\centering
0.1118

(0.0948)
\end{cellvarwidth} & \begin{cellvarwidth}[t]
\centering
-0.0023

(0.0644)
\end{cellvarwidth} & \begin{cellvarwidth}[t]
\centering
0.1234

(0.1178)
\end{cellvarwidth} & \begin{cellvarwidth}[t]
\centering
0.1672

(0.1171)
\end{cellvarwidth} & \begin{cellvarwidth}[t]
\centering
0.1027 

(0.0833)
\end{cellvarwidth} & \begin{cellvarwidth}[t]
\centering
-0.0075

(0.0725)
\end{cellvarwidth} & \begin{cellvarwidth}[t]
\centering
0.1166 

(0.0743)
\end{cellvarwidth} & \begin{cellvarwidth}[t]
\centering
0.0442

(0.0519)
\end{cellvarwidth}\tabularnewline
$\hat{\theta}(W_{RA})$  & -0.1186  & 0.1024 & -0.0163 & 0.0961  & 0.1582 & 0.1230  & -0.0376  & 0.1219  & 0.0345\tabularnewline
SE$V_{C}(W_{RA})$ & 0.0667 & 0.0700 & 0.0486 & 0.0948 & 0.0864 & 0.0655 & 0.0550 & 0.0547 & 0.0391\tabularnewline
SE$V_{Z}(W_{RA})$ & 0.0650 & 0.0676 & 0.0476 & 0.0877 & 0.0823 & 0.0621 & 0.0530 & 0.0534 & 0.0381\tabularnewline
\end{tabular}
\end{sidewaystable}

Table 2 shows the results. First, we report the original point estimates
with standard errors in parentheses. The stratified GMM point estimates
remain close to the original OLS estimates, differing by at most $0.031$
GPA points across the nine columns. Relative to the original OLS results,
the estimated variances under GMM with $V_{C}$ are $35.2\lyxmathsym{\textendash}45.8$\%
smaller. This comparison reflects the adjustment estimator, but does
not isolate the design-based variance correction. Holding each GMM
point estimate fixed, replacing $V_{C}$ with $V_{Z}$ further reduces
estimated variance by $3.8\lyxmathsym{\textendash}14.4$\%. In the
pooled-men column, this additional correction lowers the standard
error from $0.0655$ to $0.0621$ and makes the positive GMM estimate
significant at the 5\% level, whereas the original OLS estimate was
not significant.

\newpage{}

\appendix

\section{Mathematical appendix }\label{app:A}

\subsection{Proof of Theorem \ref{thm:GMM}}

\subsubsection{Proof of Part (1)}

It is sufficient to verify the conditions in Newey and McFadden (1994,
Theorem 2.1). Their condition (i) is satisfied due to the uniqueness
of $\theta^{*}$ and the positive definiteness of $W$ in Assumption
(2). Their condition (ii) (i.e., compactness of the parameter space)
is directly imposed.

To verify their conditions (iii) and (iv), it is sufficient to show
that 
\begin{equation}
\sup_{\theta\in\Theta}\left\Vert \frac{1}{N}\sum^{M}_{i=1}R_{i}g_{i}(X_{i},\theta)-\lim_{M\to\infty}\frac{1}{M}\sum^{M}_{i=1}\mathbb{E}[g_{i}(X_{i},\theta)]\right\Vert \overset{p}{\to}0,\label{pf:1-1}
\end{equation}
and $\lim_{M\to\infty}\frac{1}{M}\sum^{M}_{i=1}\mathbb{E}[g_{i}(X_{i},\theta)]$
is continuous at each $\theta\in\Theta$. The continuity of $\lim_{M\to\infty}M^{-1}\sum^{M}_{i=1}\mathbb{E}[g_{i}(X_{i},\theta)]$
follows from the Lipschitz condition in Assumption (2). For (\ref{pf:1-1}),
we first note that Abadie \emph{et al.} (2014, Lemma A.2) implies
the pointwise convergence 
\[
\frac{1}{N}\sum^{M}_{i=1}R_{i}g_{i}(X_{i},\theta)-\lim_{M\to\infty}\frac{1}{M}\sum^{M}_{i=1}\mathbb{E}[g_{i}(X_{i},\theta)]\overset{p}{\to}0,\quad\text{for each }\theta\in\Theta.
\]
Then Newey (1991, Corollary 2.2) combined with the Lipschitz condition
in Assumption (2) implies the uniform convergence in (\ref{pf:1-1}).

Since we verify all the conditions of Newey and McFadden (1994, Theorem
2.1), we obtain $\hat{\theta}_{N}(W_{N})\overset{p}{\to}\theta^{*}$.
Therefore, Assumption (2) ($\theta^{*}_{M}\to\theta^{*}$) implies
the conclusion.

\subsubsection{Proof of Part (2)}

Let $\hat{G}_{N}(\theta)=\frac{1}{N}\sum^{M}_{i=1}R_{i}\frac{\partial g_{i}(X_{i},\theta)}{\partial\theta^{\prime}}$
and $\bar{G}_{M}(\theta)=\frac{1}{M}\sum^{M}_{i=1}\mathbb{E}\left[\frac{\partial g_{i}(X_{i},\theta)}{\partial\theta^{\prime}}\right]$.
By the consistency of $\hat{\theta}_{N}(W_{N})$ and Assumption (3)
($\theta^{*}\in\mathrm{int}(\Theta)$ and differentiability of $g_{i}(x,\theta)$),
the estimator $\hat{\theta}_{N}(W_{N})$ satisfies the first-order
condition 
\[
\hat{G}_{N}(\hat{\theta}_{N}(W_{N}))^{\prime}W_{N}\left\{ \frac{1}{N}\sum^{M}_{i=1}R_{i}g_{i}(X_{i},\hat{\theta}_{N}(W_{N}))\right\} =0,
\]
with probability approaching one. By expanding $g_{i}(X_{i},\hat{\theta}_{N}(W_{N}))$
around $\theta^{*}_{M}$ and solving for $\hat{\theta}_{N}(W_{N})-\theta^{*}_{M}$,
we obtain 
\begin{equation}
\sqrt{N}(\hat{\theta}_{N}(W_{N})-\theta^{*}_{M})=-[\hat{G}_{N}(\hat{\theta}_{N}(W_{N}))^{\prime}W_{N}\hat{G}_{N}(\tilde{\theta}_{N})]^{-1}\hat{G}_{N}(\hat{\theta}_{N}(W_{N}))^{\prime}W_{N}\frac{1}{\sqrt{N}}\sum^{M}_{i=1}R_{i}g_{i}(X_{i},\theta^{*}_{M}),\label{pf:exp}
\end{equation}
where $\tilde{\theta}_{N}$ is a point on the line joining $\hat{\theta}_{N}(W_{N})$
and $\theta^{*}_{M}$. Since $\hat{\theta}_{N}(W_{N})-\theta^{*}_{M}\overset{p}{\to}0$
and $\tilde{\theta}_{N}-\theta^{*}_{M}\overset{p}{\to}0$, it is sufficient
for the conclusion to show that 
\begin{eqnarray}
\sup_{\theta\in\mathcal{N}}||\hat{G}_{N}(\theta)-\bar{G}_{M}(\theta)|| & \overset{p}{\to} & 0,\label{pf:2-1}\\
\frac{1}{\sqrt{N}}\sum^{M}_{i=1}R_{i}g_{i}(X_{i},\theta^{*}_{M}) & \overset{d}{\to} & N(0,\Omega-\rho\Delta),\label{pf:2-2}
\end{eqnarray}
for some neighborhood $\mathcal{N}$ around $\theta^{*}$.

For (\ref{pf:2-1}), Abadie \emph{et al.} (2014, Lemma A.2) and Assumption
(3) imply the pointwise convergence $\hat{G}_{N}(\theta)-\bar{G}_{M}(\theta)\overset{p}{\to}0$
for each $\theta\in\mathcal{N}$. Then Newey (1991, Corollary 2.2)
combined with the Lipschitz condition in Assumption (3) implies the
uniform convergence in (\ref{pf:2-1}). The convergence in (\ref{pf:2-2})
follows directly from Abadie \emph{et al.} (2020, Lemma A.1) under
Assumptions (1)-(3). Therefore, the conclusion follows.

\subsection{Proof of $V_{Z}((\Omega-\rho\Delta_{Z})^{-1})\le_{\mathrm{pd}}V_{Z}(\Omega^{-1})$
in Theorem \ref{thm:var} (3) }\label{app:var}

Let $A=\Omega^{-1}G(G^{\prime}\Omega^{-1}G)^{-1}$ and $B=(\Omega-\rho\Delta_{Z})^{-1}G(G^{\prime}(\Omega-\rho\Delta_{Z})^{-1}G)^{-1}$
so that $V_{Z}(\Omega^{-1})=A^{\prime}(\Omega-\rho\Delta_{Z})A$ and
$V_{Z}((\Omega-\rho\Delta_{Z})^{-1})=B^{\prime}(\Omega-\rho\Delta_{Z})B$.
Observe that 
\begin{eqnarray*}
V_{Z}(\Omega^{-1}) & = & \{B+(A-B)\}^{\prime}(\Omega-\rho\Delta_{Z})\{B+(A-B)\}\\
 & = & V_{Z}((\Omega-\rho\Delta_{Z})^{-1})+(A-B)^{\prime}(\Omega-\rho\Delta_{Z})(A-B),
\end{eqnarray*}
where the second equality follows from 
\[
B^{\prime}(\Omega-\rho\Delta_{Z})(A-B)=(G^{\prime}(\Omega-\rho\Delta_{Z})^{-1}G)^{-1}-(G^{\prime}(\Omega-\rho\Delta_{Z})^{-1}G)^{-1}=0.
\]
Since $(A-B)^{\prime}(\Omega-\rho\Delta_{Z})(A-B)\ge_{\mathrm{pd}}0$,
we obtain the conclusion.

\subsection{Proof of Corollary \ref{cor:mis}}

\subsubsection{Proof of Part (1)}

Let $\hat{Q}_{N}(\theta,W_{N})=\hat{g}_{N}(\theta)'W_{N}\hat{g}_{N}(\theta)$.
The uniform law of large numbers established in the proof of Theorem
\ref{thm:GMM} implies 
\[
\sup_{\theta\in\Theta}|\hat{Q}_{N}(\theta,W_{N})-Q_{M}(\theta,W)|\overset{p}{\to}0.
\]
Since $\theta^{\#}_{M}(W)$ uniquely minimizes $Q_{M}(\theta,W)$,
the result follows from the standard extremum-estimator argument of
Newey and McFadden (1994, Theorem 2.1).

\subsubsection{Proof of Parts (2) and (3)}

Suppress the dependence on $W$ and write $\theta^{\#}_{M}=\theta^{\#}_{M}(W)$,
$\hat{g}_{N}(\theta)=N^{-1}\sum_{i}R_{i}g_{i}(X_{i},\theta)$, and
$\hat{G}_{N}(\theta)=N^{-1}\sum_{i}R_{i}D_{i}(X_{i},\theta)$. The
population first-order condition is 
\[
G^{\#\prime}_{M}W\mu^{\#}_{M}=0.
\]
For a fixed weight, expand the sample score $S_{N}(\theta,W)=\hat{G}_{N}(\theta)^{\prime}W\hat{g}_{N}(\theta)$
around $\theta^{\#}_{M}$. Uniform convergence of the first two derivatives
gives 
\[
0=S_{N}(\theta^{\#}_{M},W)+\mathcal{\hat{H}}_{N}(\tilde{\theta}_{N})(\widehat{\theta}_{N}-\theta^{\#}_{M}),\qquad\mathcal{\hat{H}}_{N}(\widetilde{\theta}_{N})\overset{p}{\to}\mathcal{H}^{\#}.
\]
The key centering step is 
\begin{eqnarray*}
\sqrt{N}S_{N}(\theta^{\#}_{M},W) & = & G^{\#\prime}_{M}W\sqrt{N}\{\widehat{g}_{N}(\theta^{\#}_{M})-\mu^{\#}_{M}\}+\sqrt{N}\{\widehat{G}_{N}(\theta^{\#}_{M})-G^{\#}_{M}\}^{\prime}W\mu^{\#}_{M}+o_{p}(1)\\
 & = & \frac{1}{\sqrt{N}}\sum^{M}_{i=1}R_{i}\psi^{\#}_{iM}+o_{p}(1),
\end{eqnarray*}
where the last equality uses $N^{-1}\sum_{i}R_{i}=1$. Since $M^{-1}\sum_{i}\mathbb{E}[\psi^{\#}_{iM}]=0$,
the finite-population central limit theorem gives 
\[
\frac{1}{\sqrt{N}}\sum^{M}_{i=1}R_{i}\psi^{\#}_{iM}\overset{d}{\to}N(0,\Omega^{\#}_{\psi}-\rho\Delta^{\#}_{\psi}).
\]
Solving the score expansion yields 
\[
\sqrt{N}(\hat{\theta}_{N}-\theta^{\#}_{M})=-(\mathcal{H}^{\#})^{-1}\frac{1}{\sqrt{N}}\sum_{i}R_{i}\psi^{\#}_{iM}+o_{p}(1),
\]
which proves Part (2). If $W_{N}$ is estimated, the additional first-order
term is 
\[
G^{\#\prime}_{M}\sqrt{N}(W_{N}-W)\mu^{\#}_{M}+o_{p}(1).
\]
Joint convergence and Slutsky's theorem then yield Part (3).

\subsection{Proof of Corollary \ref{cor:overid}}

Let $\hat{g}_{N}(\theta^{*}_{M})=\frac{1}{N}\sum^{M}_{i=1}R_{i}g_{i}(X_{i},\theta^{*}_{M})$
and $\hat{G}_{N}(\theta^{*}_{M})=\frac{1}{N}\sum^{M}_{i=1}R_{i}\frac{\partial g_{i}(X_{i},\theta^{*}_{M})}{\partial\theta^{\prime}}$.
Assumptions (1)--(3) imply $\hat{G}_{N}(\theta^{*}_{M})\overset{p}{\to}G$,
$W_{N}\overset{p}{\to}W$, and 
\[
\sqrt{N}\hat{g}_{N}(\theta^{*}_{M})\overset{d}{\to}Z\sim N(0,\Omega-\rho\Delta).
\]
Thus, an expansion of the first-order condition of $\hat{\theta}_{N}$
yields 
\begin{eqnarray*}
\sqrt{N}(\hat{\theta}_{N}-\theta^{*}_{M}) & = & -\{\hat{G}_{N}(\hat{\theta}_{N})^{\prime}W_{N}\hat{G}_{N}(\bar{\theta})\}^{-1}\hat{G}_{N}(\hat{\theta}_{N})^{\prime}W_{N}\sqrt{N}\hat{g}_{N}(\theta^{*}_{M})+o_{p}(1)\\
 & \overset{d}{\to} & -(G^{\prime}WG)^{-1}G^{\prime}WZ.
\end{eqnarray*}
Combining these results, we obtain 
\begin{eqnarray*}
\sqrt{N}\hat{g}_{N}(\hat{\theta}_{N}) & = & [I_{k}-\hat{G}_{N}(\bar{\theta})\{\hat{G}_{N}(\hat{\theta}_{N})^{\prime}W_{N}\hat{G}_{N}(\bar{\theta})\}^{-1}\hat{G}_{N}(\hat{\theta}_{N})^{\prime}W_{N}]\sqrt{N}\hat{g}_{N}(\theta^{*}_{M})+o_{p}(1)\\
 & \overset{d}{\to} & \{I_{k}-G(G^{\prime}WG)^{-1}G^{\prime}W\}Z.
\end{eqnarray*}
Therefore, the test statistic satisfies 
\[
J_{N}(W_{N})\overset{d}{\to}Z^{\prime}\{W-WG(G^{\prime}WG)^{-1}G^{\prime}W\}Z=:Q(W).
\]
For Part (1) of this corollary, let $W=(\Omega-\rho\Delta)^{-1}$
and $\tilde{Z}=(\Omega-\rho\Delta)^{-1/2}Z\sim N(0,I_{k})$. Then
we have $Q(W)=\tilde{Z}^{\prime}(I_{k}-P)\tilde{Z}$, where 
\[
P=(\Omega-\rho\Delta)^{-1/2}G(G^{\prime}(\Omega-\rho\Delta)^{-1}G)^{-1}G^{\prime}(\Omega-\rho\Delta)^{-1/2}.
\]
Since $I_{k}-P$ is the symmetric idempotent matrix of rank $k-p$,
we obtain the conclusion $Q(W)\sim\chi^{2}_{k-p}$.

For Part (2) of this corollary, Hansen (2021, Lemma 1) implies that
if $\{\lambda_{j}\}^{k-p}_{j=1}$ are the nonzero eigenvalues of $(\Omega-\rho\Delta)\{W-WG(G^{\prime}WG)^{-1}G^{\prime}W\}$,
then $Q(W)\sim\sum^{k-p}_{j=1}\lambda_{j}\chi^{2}_{1,j}$. To show
the eigenvalue ordering used in Section \ref{sub:overid}, let $A=W-WG(G^{\prime}WG)^{-1}G^{\prime}W$.
Since $A$ is positive semidefinite and $\Delta_{Z}\leq_{\mathrm{pd}}\Delta$,
we have $A^{1/2}(\Omega-\rho\Delta_{Z})A^{1/2}\geq_{\mathrm{pd}}A^{1/2}(\Omega-\rho\Delta)A^{1/2}$.
The nonzero eigenvalues of these two matrices are the same as those
of $(\Omega-\rho\Delta_{Z})A$ and $(\Omega-\rho\Delta)A$, respectively.
Thus, ordering the nonzero eigenvalues in decreasing order, the Courant-Fischer-Weyl
min-max principle implies $\lambda_{Z,j}\geq\lambda_{j}$ for $j=1,\ldots,k-p$.

\section{Additional tables}

\begin{sidewaystable}[H]
\vspace{10cm}

\caption{Simulation results ($K=5$)}

\begin{tabular}{c|ccc|ccc|ccc}
\hline 
 & \multicolumn{3}{c|}{$M=500$} & \multicolumn{3}{c|}{$M=1000$} & \multicolumn{3}{c}{$M=2000$}\tabularnewline
 & $\rho=0.25$ & $\rho=0.5$ & $\rho=0.75$ & $\rho=0.25$ & $\rho=0.5$ & $\rho=0.75$ & $\rho=0.25$ & $\rho=0.5$ & $\rho=0.75$\tabularnewline
\hline 
$\lvert\bar{\hat{\beta}}(\hat{\Omega}^{-1})-\beta^{*}_{M}\rvert$ & 0.0006 & 0.0014 & 0.0008 & 0.0009 & 0.0002 & 0.0003 & 0.0000 & 0.0000 & 0.0003\tabularnewline
$\lvert\bar{\hat{\beta}}((\hat{\Omega}-\rho\hat{\Delta}_{Z})^{-1})-\beta^{*}_{M}\rvert$ & 0.0002 & 0.0007 & 0.0004 & 0.0007 & 0.0002 & 0.0002 & 0.0001 & 0.0000 & 0.0001\tabularnewline
\hline 
SD $\hat{\beta}(\hat{\Omega}^{-1})$ & 0.0923 & 0.0591 & 0.0412 & 0.0670 & 0.0428 & 0.0298 & 0.0484 & 0.0296 & 0.0204\tabularnewline
SD $\hat{\beta}((\hat{\Omega}-\rho\hat{\Delta}_{Z})^{-1})$ & 0.0914 & 0.0563 & 0.0350 & 0.0660 & 0.0404 & 0.0244 & 0.0480 & 0.0283 & 0.0171\tabularnewline
$\sqrt{V_{\mathrm{GMM}}(\Omega^{-1})/N}$ & 0.0958 & 0.0599 & 0.0415 & 0.0677 & 0.0423 & 0.0293 & 0.0479 & 0.0299 & 0.0208\tabularnewline
$\sqrt{V_{\mathrm{GMM}}((\Omega-\rho\Delta_{Z})^{-1})/N}$ & 0.0950 & 0.0572 & 0.0347 & 0.0672 & 0.0404 & 0.0246 & 0.0475 & 0.0286 & 0.0174\tabularnewline
\hline 
SE $V_{\mathrm{C}}(\Omega^{-1})$ & 0.0959 & 0.0710 & 0.0590 & 0.0710 & 0.0514 & 0.0424 & 0.0515 & 0.0368 & 0.0302\tabularnewline
SE $V_{Z}(\Omega^{-1})$ & 0.0884 & 0.0584 & 0.0416 & 0.0650 & 0.0418 & 0.0295 & 0.0469 & 0.0298 & 0.0209\tabularnewline
SE $V_{Z}((\Omega-\rho\Delta_{Z})^{-1})$ & 0.0877 & 0.0558 & 0.0350 & 0.0645 & 0.0400 & 0.0248 & 0.0465 & 0.0285 & 0.0175\tabularnewline
\hline 
Coverage $V_{\mathrm{C}}(\Omega^{-1})$ & 0.9668 & 0.9796 & 0.9942 & 0.9636 & 0.9804 & 0.9928 & 0.9610 & 0.9866 & 0.9968\tabularnewline
Coverage $V_{Z}(\Omega^{-1})$ & 0.9460 & 0.9514 & 0.9550 & 0.9440 & 0.9434 & 0.9482 & 0.9434 & 0.9508 & 0.9550\tabularnewline
Coverage $V_{Z}((\Omega-\rho\Delta_{Z})^{-1})$ & 0.9472 & 0.9496 & 0.9478 & 0.9432 & 0.9436 & 0.9534 & 0.9434 & 0.9498 & 0.9538\tabularnewline
\hline 
$s^{*}_{\ell}$ & 0.5111 & 0.5886 & 0.7022 & 0.5251 & 0.6010 & 0.7152 & 0.5328 & 0.6074 & 0.7218\tabularnewline
\hline 
Chi-Squared & 0.0372 & 0.0422 & 0.0358 & 0.0430 & 0.0406 & 0.0386 & 0.0448 & 0.0434 & 0.0432\tabularnewline
Weighted Chi-Squared & 0.0428 & 0.0508 & 0.0466 & 0.0484 & 0.0496 & 0.0474 & 0.0504 & 0.0494 & 0.0540\tabularnewline
\end{tabular}
\end{sidewaystable}

\end{document}